\documentclass{article}

\pdfoutput=1

\usepackage{authblk}
\usepackage{graphicx}
\usepackage{caption}
\usepackage{subcaption}
\usepackage{bm}
\usepackage{verbatim}
\usepackage{eqnarray}
\usepackage{changepage}
\usepackage{amsmath, amsthm, amssymb}
\usepackage[top=1.1in, bottom=1.15in, left=1in, right=1in]{geometry}

\newcommand*\diff{\mathop{}\!\mathrm{d}}

\DeclareMathOperator\erf{erf}

\begin{document}

\begin{flushleft}
	{\Large
		\textbf\newline{Intrinsically-generated fluctuating activity in excitatory-inhibitory networks}
	}
	\newline
	\\
	Francesca Mastrogiuseppe\textsuperscript{1,2, *},
	Srdjan Ostojic\textsuperscript{1, *}
	\\
	\bigskip
	\bf{1} Laboratoire de Neurosciences Cognitives, INSERM U960 and
	\\
	\bf{2} Laboratoire de Physique Statistique, CNRS UMR 8550. \'Ecole Normale Sup\'erieure - PSL Research University, Paris, France
	\\
	\bigskip
	
	% Use the asterisk to denote corresponding authorship and provide email address in note below.
	* francesca.mastrogiuseppe@ens.fr (FM), srdjan.ostojic@ens.fr (SO)
	
\end{flushleft}

\section*{Abstract}
Recurrent networks of non-linear  units display a variety of dynamical
regimes depending  on the structure of their  synaptic connectivity. A
particularly  remarkable  phenomenon  is  the appearance  of  strongly
fluctuating,  chaotic  activity  in  networks  of  deterministic,  but
randomly  connected  rate  units.   How  this  type  of  intrinsically
generated fluctuations  appears in more realistic  networks of spiking
neurons has  been a  long standing question.   To ease  the comparison
between  rate  and spiking  networks,  recent  works investigated  the
dynamical regimes of  randomly-connected rate networks with segregated
excitatory and inhibitory populations, and firing rates constrained to
be positive.   These works derived general dynamical  mean field (DMF)
equations  describing  the  fluctuating  dynamics,  but  solved  these
equations only  in the  case of purely  inhibitory networks.   Using a
simplified excitatory-inhibitory  architecture in which  DMF equations
are  more  easily  tractable,  here  we  show  that  the  presence  of
excitation qualitatively modifies the fluctuating activity compared to
purely inhibitory networks.   In presence of excitation, intrinsically
generated  fluctuations induce  a strong increase  in mean  firing  rates, a
phenomenon that is much weaker in  purely inhibitory networks.  Excitation moreover
induces  two  different  fluctuating  regimes:  for  moderate  overall
coupling,   recurrent   inhibition    is   sufficient   to   stabilize
fluctuations; for strong coupling,  firing rates are stabilized solely
by the upper bound imposed on activity, even if inhibition is stronger
than  excitation.   These  results  extend  to  more  general  network
architectures, and  to rate networks receiving  noisy inputs mimicking
spiking  activity.  Finally,  we show  that signatures  of  the second
dynamical regime appear in networks of integrate-and-fire neurons.

\section*{Author Summary}
Electrophysiological recordings from cortical circuits reveal strongly
irregular  and  highly complex  temporal  patterns  of in-vivo  neural
activity. In the  last decades, a large number  of theoretical studies
have  speculated on  the possible  sources of  fluctuations  in neural
assemblies,   pointing   out   the   possibility   of   self-sustained
irregularity,  intrinsically  generated  by  network  mechanisms.   In
particular,  a seminal  study  showed that  purely deterministic,  but
randomly  connected   rate  networks  intrinsically   develop  chaotic
fluctuations due to the recurrent  feedback.  In the simple and highly
symmetric  class   of  models  considered  in   classical  works,  the
transition from  stationary activity to chaos is  characterized by the
behavior  of the  auto-correlation function  and the  critical slowing
down of  fluctuations. Following up  on recent works, here  we combine
analytical and numerical tools to investigate the macroscopic dynamics
generated by  more realistic models of excitatory  and inhibitory rate
units.   We show  that  the presence  of  excitation leads  to a  strong
signature of the  onset of chaos in the  first-order statistics of the
network activity, and  that this effect is highly  robust with respect
to  spiking noise.   We moreover  find  that excitation  leads to  two
different types of fluctuating  activity at moderate and strong
synaptic coupling, even when inhibition dominates.  Finally, we test the
appearance   of   analogous   dynamical   regimes   in   networks   of
integrate-and-fire neurons.

\section*{Introduction}
Networks  of   excitatory  and  inhibitory  neurons   form  the  basic
processing   units  in  the   cortex.   Understanding   the  dynamical
repertoire of  such networks is therefore  essential for understanding
their input-output  properties and identifying  potential computational
mechanisms in the brain.

One  of the  simplest models  of a  cortical network  is a  network of
randomly connected units, the  activity of each unit being represented
by its instantaneous firing rate.   A seminal study revealed that such
networks can exhibit a  transition from constant to strongly irregular
activity when the coupling  is increased \cite{Sompolinsky}. Above the
transition, the  network displays  a state in  which the  firing rates
fluctuate strongly in time and across units, although the dynamics are
fully deterministic and there are no external inputs.  Such internally
generated fluctuating activity is a signature of the chaotic nature
of the dynamics  \cite{Molgedey,Cessac,Cessac2}, and the corresponding
regime  has been referred  to as  rate chaos.   Recently, it  has been
proposed  that this  type of  activity can  serve as  a  substrate for
complex computations \cite{BuonomanoMaass}.  Several works showed that
the randomly connected rate network  is able to learn complex temporal
dynamics            and            input-output           associations
\cite{Sussillo,Buonomano,Sussillo2}.   These  computational properties
may be related to the  appearance of an exponential number of unstable
fixed points  at the transition  \cite{Wainrib}, and to the  fact that
dynamics   are   slow    and   the   signal-to-noise   ratio   maximal
\cite{Toyoizumi}.

A  natural question  is  whether actual  cortical  networks exhibit  a
dynamical regime analogous to rate chaos \cite{Ostojic}.  The classical network model
analyzed     in    \cite{Sompolinsky}    and     subsequent    studies
\cite{Sussillo,Buonomano,Rajan,   Aljadeff1,Aljadeff2,Stern,Goedeke2016} contains
several  simplifying features  that prevent  a direct  comparison with
more  biologically  constrained models  such  as  networks of  spiking
neurons. In  particular, a  major simplification is  a high  degree of
symmetry  in both  input currents  and firing  rates.  Indeed,  in the
classical model  the synaptic strengths  are symmetrically distributed
around zero, and excitatory  and inhibitory neurons are not segregated
into   different  populations,   thus  violating   Dale's   law.   The
current-to-rate  activation function  is furthermore  symmetric around
zero, so  that the dynamics are  symmetric under sign  reversal.  As a
consequence, the mean activity in  the network is always zero, and the
transition to the fluctuating  regime is characterized solely in terms
of  second  order  statistics.

To help bridge the gap  between the classical model and more realistic
spiking  networks   \cite{Brunel2000,  Ostojic},  recent   works  have
investigated  fluctuating  activity  in  rate  networks  that  include
additional  biological constraints  \cite{Ostojic,Kadmon,Harish}, such
as  segregated excitatory-and-inhibitory populations,  positive firing
rates  and spiking noise  \cite{Kadmon}. In  particular, two  of those
works \cite{Kadmon,Harish}  extended to excitatory-inhibitory networks
the dynamical mean field (DMF)  theory used for the  analysis of rate
chaos   in    classical   works   \cite{Sompolinsky}.     In   general
excitatory-inhibitory  networks,  the  DMF equations  however  proved
difficult to solve, and these works focused instead mostly on the case
of  purely inhibitory networks. These works therefore  left unexplained  some phenomena observed in    simulations   of   excitatory-inhibitory    spiking and rate  networks
\cite{Ostojic,Engelken,Ostojic2015}, in particular the observation that the onset of fluctuating activity is accompanied by a large elevation of mean firing rate \cite{Ostojic}, and the finding that fluctuating activity at strong coupling is highly sensitive to the upper bound \cite{Ostojic2015}.

Here we investigate the  effects of excitation on fluctuating activity
in inhibition-dominated excitatory-inhibitory networks \cite{Troyer97,
	Miller1, Ahmadian,  Miller2, Hennequin2012, Hennequin2014}.  To this  end, we focus
on  a   simplified  network  architecture  in   which  excitatory  and
inhibitory    neurons   receive    statistically    identical   inputs
\cite{Brunel2000}.   For  that   architecture,   dynamical  mean field
equations can be  solved. We find that in presence of excitation,
the  coupling between mean  and the  auto-correlation of  the activity
leads to  a strong increase of mean  firing rates in  the fluctuating regime
\cite{Ostojic},  a phenomenon  that is much weaker in  purely  inhibitory networks.
Moreover,  as the  coupling  is increased,  two  different regimes  of
fluctuating   activity   appear:   at   intermediate   coupling,   the
fluctuations are  of moderate amplitude and  stabilized by inhibition;
at  strong  coupling, the  fluctuations  become  very  large, and  are
stabilized only by an upper  bound on the activity, even if inhibition
globally dominates. The second regime  is highly robust to external or
spiking noise, and appears also in more general network architectures.  Finally we show that networks
of  spiking   neurons  exhibit  signatures   characteristic  of  these
different regimes.  

\section*{Results}

We consider a large, randomly connected network of excitatory and inhibitory rate units  similar to previous studies \cite{Kadmon,Harish}. The network dynamics are given by:
\begin{equation}
	\dot{x_i}(t)=-x_i(t)+\sum_{j=1}^{N} J_{ij} \phi (x_j(t))+I
	\label{eq:dyn}
\end{equation}
where $N$  is the  total number of  units, $x_i$ represents  the total
input current  to unit $i$, and  $J_{ij}$ is the strength of  synaptic inputs
from unit $j$ to unit $i$. In most of the results which follow, we will not include any external currents ($I=0$).  The
function  $\phi(x)$  is  a  monotonic, positively  defined  activation
function  that  transforms input  currents  into  output activity.  For
mathematical  convenience,   in  most  of   the  analysis  we   use  a
threshold-linear activation with an upper-bound $\phi_{max}$ (see \emph{Methods}).

We  focus on a  sparse,  two-population synaptic  matrix identical  to
\cite{Brunel2000,Ostojic}. We  first study the simplest  version in  which all neurons receive the same  number $C \ll N$
of  incoming  connections   (respectively  $C_E=fC$  and  $C_I=(1-f)C$
excitatory and  inhibitory inputs).
All the  excitatory synapses have
strength $J$ and all inhibitory synapses have strength $-gJ$, but the
precise  pattern  of  connections  is  assigned  randomly.   For  such
connectivity,  excitatory  and  inhibitory neurons  are  statistically
equivalent  as  they  receive  statistically identical  inputs.   This
situation greatly simplifies the  mathematical analysis, and allows us
to obtain results  in a transparent manner. In a  second step, we show
that   the  obtained  results   extend  to   more  general   types  of
connectivity.

\subsection*{Emergence of fluctuations in deterministic networks}

\paragraph*{Dynamical systems analysis}
For  a fixed,  randomly  chosen connectivity  matrix,  the network  we
consider  is fully  deterministic,  and can  therefore   be  examined 
in a  first approach   using  standard  dynamical  system  techniques
\cite{Strogatz}.  Such an analysis has been performed in a number of previous studies (see e.g. \cite{Ostojic,Hennequin2014}), here we include it for completeness.

As the  inputs to all  units are statistically identical,  the network
admits a homogeneous fixed point  in which the activity is constant
in time and identical for all units, given by:
\begin{equation}
	x_0=J(C_E-gC_I)\phi(x_0)+I. 
	\label{eq:fp}
\end{equation}
The  linear  stability  of  this  fixed point  is  determined  by  the
eigenvalues of the   matrix $S_{ij}=\phi'(x_0)J_{ij}$. If the
real parts of all eigenvalues are smaller than one, the fixed point is
stable, otherwise it is linearly unstable.

For  large  networks, the  eigenspectrum  of  $J_{ij}$  consists of  a
part that  is densely  distributed in  the
complex plane  over a  circle of  radius $J\sqrt{C_E+g^2 C_I}$, and of  a real outlier
given   by  the   effective  balance of   excitation and inhibition   in  the
connectivity $J(C_E-gC_I)$  \cite{Rajan2,Tao,Tao2}.  We focus  here on
an inhibition-dominated network corresponding to $g>C_E/C_I$.  In this
regime, the real  outlier is always negative and  the stability of the
fixed   point   depends  only   on   the   densely distributed   part  of   the
eigenspectrum. The  radius of the  eigenspectrum disk, in
particular, increases with the  coupling $J$, and an instability occurs
when the radius  crosses unity.  The critical coupling  $J_0$ is given
by:
\begin{equation}
	\phi'(x_0)J_0\sqrt{C_E+g^2C_I}=1
\end{equation}
where $x_0$ depends implicitly on $J$ through Eq.~\eqref{eq:fp} and the gain  $\phi'(x)$ is in general finite
and non-negative for all the values of $x$.

\begin{figure}[h!]
	\begin{adjustwidth}{-0.3in}{-0.3in} 
		\centering
		\includegraphics{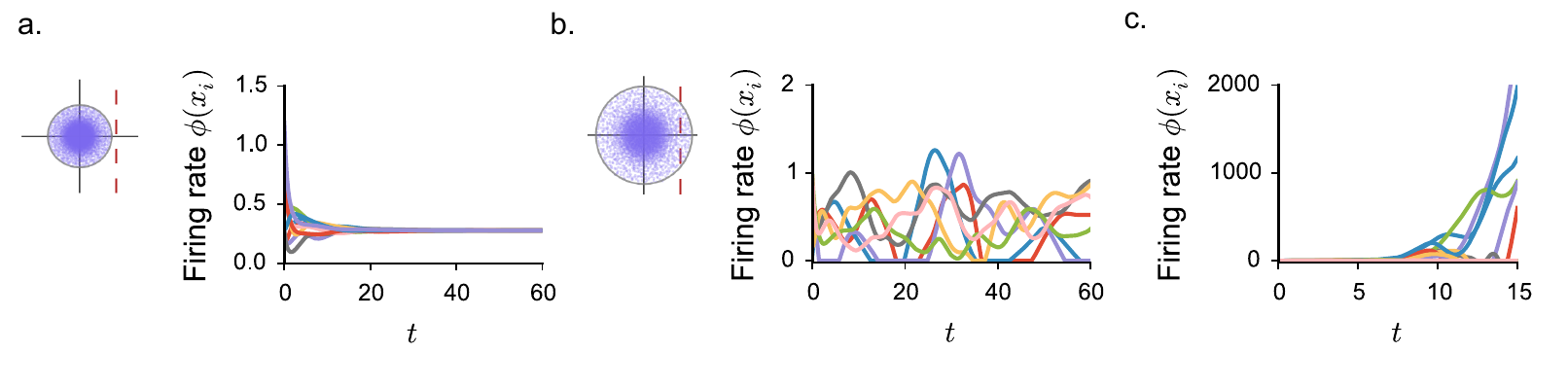}
		\caption{ {\bf Dynamical regimes of an excitatory-inhibitory network of threshold-linear units as the coupling is increased.}
			Numerical integration of the dynamics in Eq.~\eqref{eq:dyn}, firing rates of randomly chosen units. In the insets: complex eigenspectrum of the fixed point stability matrix, the red line corresponding to the stability bound.  \textbf{a.}  Weak coupling regime: the network activity converges to the homogeneous fixed point. \textbf{b.} Intermediate coupling regime: the activity displays stable fluctuations in time and across different units. \textbf{c.} Strong coupling regime: in absence of an upper bound, activity diverges.
			Choice of the parameters: $g=4.5$, $C=100.$ $N=2000 $, no saturating upper bound: $\phi_{max}\rightarrow \infty$. In this and all other figures, all quantities are unitless (see Methods). }
		\label{fig:act}
	\end{adjustwidth}
\end{figure}

Numerical  simulations confirm  that, when  $J<J_0$,  network activity
settles  into the  homogeneous  fixed point  given by  Eq.~\eqref{eq:fp}
(Fig.~\ref{fig:act} {\bf a}).  For $J>J_0$,  the fixed point is unstable,
and  the network  exhibits  ongoing   dynamics  in which  the
activities of different neurons fluctuate irregularly both in time and
across units (Fig.~\ref{fig:act} {\bf b}). As the system is deterministic,
these  fluctuations  are generated  intrinsically  in  the network  by
strong feedback  along unstable modes,  which possess a random structure
inherited from the random connectivity matrix.

\paragraph*{Dynamical mean field description}
The irregular,  fluctuating activity regime cannot  be easily analyzed
with the tools of classical dynamical systems.  Rather than attempting
to describe  single trajectories, we  follow a different  approach and
focus on their statistics determined by averaging over time, instances
of the  connectivity matrix  and initial conditions.  To this  end, we
exploit mean field methods initially introduced for stochastic systems
consisting   of  large  numbers   of  units   \cite{Zippelius}.   More
specifically,  we  apply  to  our specific  network  architecture  the
dynamical  mean  field   approach  previously  developed  for  similar
deterministic                                                  networks
\cite{Sompolinsky,Rajan,Molgedey,Kadmon,Harish}.

Dynamical Mean Field  (DMF)  acts  by  replacing  the  fully
deterministic interacting network  by an equivalent stochastic system.
As the interaction  between units $\sum_{j} J_{ij} \phi
(x_j)$ consists of a sum of a large number of terms, it can be replaced by
a  Gaussian  stochastic   process  $\eta_i(t)$.   Such  a  replacement
provides an exact  mathematical description under specific assumptions
on  the  chaotic nature  of  the  dynamics  \cite{BenArous}, and  for
particular limits of large network  size $N$ and number of connections
$C$.  Here  we will treat it  as an approximation, and  we will assess
the  accuracy of  this  approximation by  comparing  the results  with
simulations  performed  for  fixed  $C$  and $N$  (see  
\emph{Methods} for the limits of this approximation).

Replacing the  interaction terms by Gaussian  processes transforms the
system into $N$ identical Langevin-like equations:
\begin{equation}
	\dot{x_i}(t)=-x_i(t)+\eta_i(t).
	\label{eq:langevin}
\end{equation}
As $\eta_i(t)$  is a Gaussian noise, each  trajectory $x_i(t)$ emerges
thus as a Gaussian stochastic process, characterized by its first- and
second-order moments.   Within DMF, the mean and  correlations of this
stochastic  process  are  determined self-consistently,  by  replacing
averages  over $\eta_i$  with  averages over  time,  instances of  the
connectivity matrix and initial conditions in the original system.  In
the limit  of a large network, the  stochastic processes corresponding
to different units become uncorrelated. Moreover, in the
specific  network  architecture   considered  here,  all  units  are
statistically  equivalent,  so   that  the  network  is  effectively
described  by   a  single  process.   Note  that  in   more  general
excitatory-inhibitory  networks,  a  distinction  needs to  be  made
between  different  classes  of  neurons, and  the  DMF  description
becomes more complex \cite{Kadmon,Harish}.  The details of the mean
field analysis are provided in \emph{Methods}.

The final outcome  of DMF is a set of two equations for the first- and second-order statistics of the network activity.
The equations are written in terms of  the mean
$[\phi]$ and autocorrelation $C(\tau)$
of   the  firing   rate  and   the  mean   $\mu$   and  mean-subtracted autocorrelation
$\Delta(\tau)$ of the input currents. The two sets of statistics provide an equivalent description of activity and have to respect self-consistency:
\begin{align}
	& [\phi]=\int \mathcal{D}z \phi(\mu + \sqrt{\Delta_0}z) 
	\label{eq:dmf-mean1}\\
	& C(\tau)=\int \mathcal{D}z \left[ \int \mathcal{D}y \phi(\mu +\sqrt{\Delta_0-|\Delta(\tau)|} y + \sqrt{|\Delta(\tau)|}z) \right]^2
	\label{eq:dmf-mean2}
\end{align}

where
\begin{align}
	&\mu= J(C_E-gC_I)[\phi]+I  \label{eq:dmf-mu}\\
	& C(\tau)=[\phi(x_i(t))\phi(x_i(t+\tau))] \\
	& \Delta(\tau)=[x_i(t)x_i(t+\tau)]-[x_i]^2.
\end{align}

In Eqs.~\eqref{eq:dmf-mean1}-\eqref{eq:dmf-mean2}  we   used  the   short-hand  notation:  $\int   \mathcal{D}z  =
\int_{-\infty}^{+\infty}  \frac{e^{-\frac{z^2}{2}}}{\sqrt{2 \pi}} dz$,
and  $\Delta_0=\Delta(\tau=0)$.  Note that  since  all  the units  are
statistically  equivalent, $[\phi]$ and  $C(\tau)$ are  independent of
the index $i$. The input current correlation function $\Delta(\tau)$ moreover obeys an evolution equation in which the mean $[\phi]$ enters:
\begin{equation}
	\ddot{\Delta}(\tau)=\Delta(\tau)-J^2(C_E+g^2C_I)\{C(\tau)-[\phi]^2\}.
	\label{eq:deltaev}
\end{equation} 

The   main   difference  here    with   respect   to   classical   works
\cite{Sompolinsky} is  that the first-order statistics  are not trivial.
In the classical  case, the mean input $\mu$  is zero by construction,
and  the activation function $\phi(x)=\tanh(x)$ is  symmetric  around zero,  so  that the  mean
firing rate $[\phi]$ in Eq.~\eqref{eq:dmf-mean1} is zero.  In our case,
firing-rates are constrained to be  positive, so that even in the case
of  perfect  balance  ($\mu=0$),  the  mean firing  rate  $[\phi]$  can
in general be positive.   We stress that as a  consequence, the
dynamics  are  described  by  coupled  equations for  the  first-  and
second-order statistics rather than by second-order statistics alone  (see also \cite{Kadmon,Harish}).

\begin{figure}[h!]
	\begin{adjustwidth}{-0.3in}{-0.3in} 
		\centering
		\includegraphics{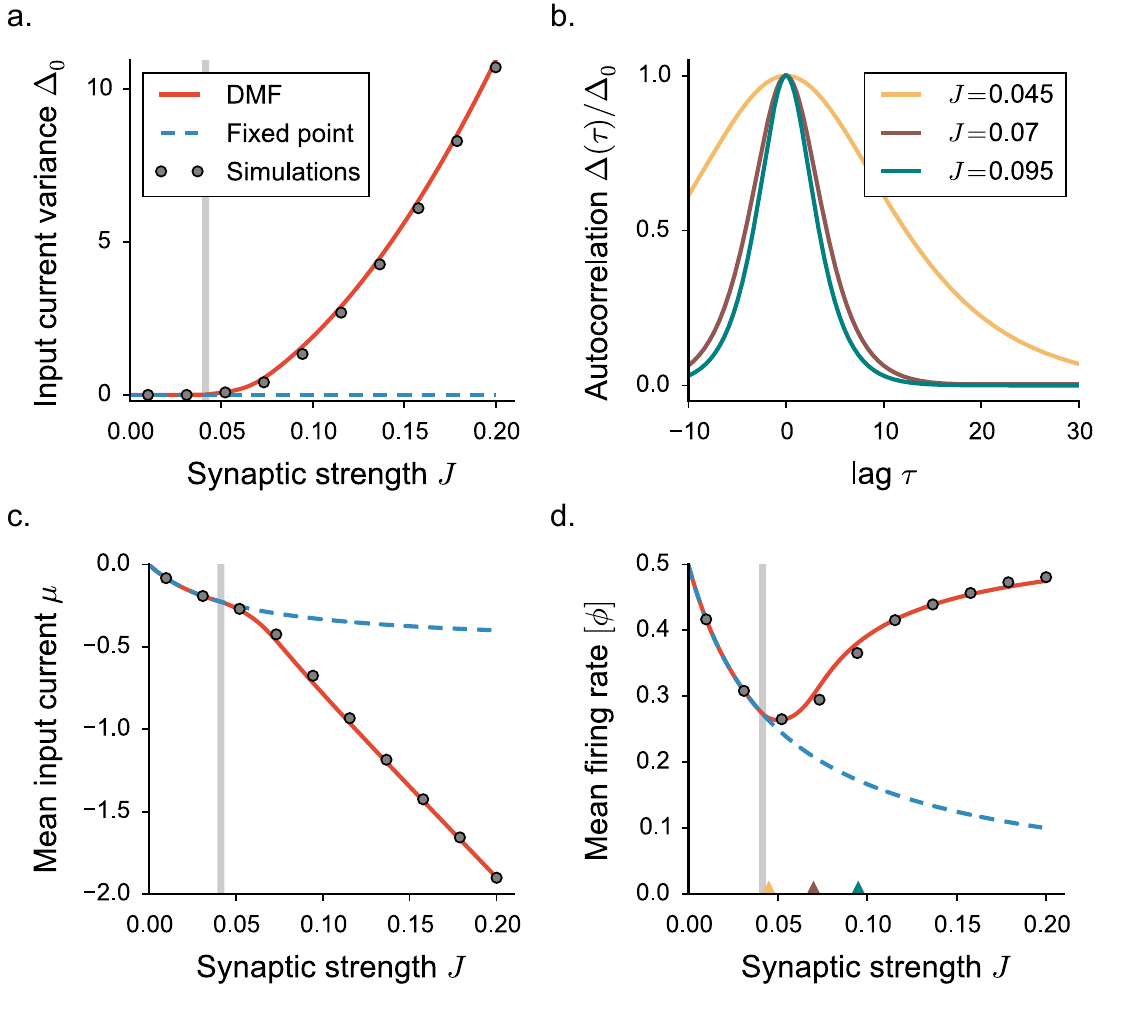}
		\caption{ {\bf Statistical description of the network activity with a threshold-linear activation function.}
			The dynamics mean field results are shown in full lines, numerical simulations as points.  \textbf{a.} Input current variance as a function of the synaptic coupling $J$. Vertical grey lines indicate the critical value $J_C$. Grey points show time and population averages performed on 4 realizations of simulated networks, $N=7000$. \textbf{b.} Normalized auto-correlation function for increasing values of the synaptic coupling (indicated by colored triangles in panel \textbf{d}). \textbf{c-d.} First order statistics: mean input current and mean firing rate. Choice of the parameters: $g=5$, $C=100$, $\phi_{max}=2$. }
		\label{fig:three}
	\end{adjustwidth}
\end{figure}

Because all units are statistically equivalent, the DMF equations can be solved, and yield
for each  set of network parameters the  mean-firing rate $[\phi]$,
the  mean  input  current  $\mu$,  the  current variance
$\Delta_0$     and     the     current    correlation     function
$\Delta(\tau)$.  Fig.~\ref{fig:three} shows  a good  match between
theoretical  predictions  and  numerically simulated  activity.  A  more  detailed
analysis of finite size effects and limitations in DMF can be found in
the \emph{Methods}.

In  agreement with  the dynamical systems analysis,  for  low coupling
values, DMF predicts a solution  for which the variance $\Delta_0$ and
the autocorrelation  $\Delta(\tau)$ of the fluctuations  vanish at all
times.   Input  currents set  into  a  stationary  and uniform  value,
corresponding  to their  mean  $\mu$.  The  predicted  value of  $\mu$
coincides with  the fixed point $x_0$, representing  a low firing-rate
background  activity.  As the  coupling  $J$  is  increased, the  mean
current  becomes increasingly  negative because  inhibition dominates,
and the mean firing rate decreases (Fig.~\ref{fig:three} {\bf c-d}).

For a critical coupling  strength $J=J_C$ (which coincides with $J_0$,
where  the fixed  point solution  loses stability),  DMF  predicts the
onset  of a  second  solution with   fluctuations of non-vanishing magnitude.  Above
$J_C$,  the   variance  of  the  activity  grows   smoothly  from  $0$
(Fig.~\ref{fig:three} {\bf a}),   and  the   auto-correlation  $\Delta(\tau)$
acquires a temporal structure, exponentially decaying to zero as $\tau
\rightarrow  \infty$. Close  to  the critical  coupling, the  dynamics
exhibit a  critical slowing down  and the decay timescale  diverges at
$J_C$,  a  behavior  characteristic  of a  critical  phase  transition
\cite{Sompolinsky} (Fig.~\ref{fig:three} {\bf b}).

The  onset of irregular,  fluctuating activity  is characterized  by a
transition of the second-order statistics from zero to a non-vanishing
value. The appearance of  fluctuations, however, directly affects also
the first-order  statistics.  As  the  firing  rates  are  constrained  to  be
positive, large fluctuations induce deviations of the mean firing rate
$[\phi]$  and the  mean input  current  $\mu$ from  their fixed  point
solutions.   In  particular,  as  $J$  increases,  larger  and  larger
fluctuations in the current lead  to an effective increase in the mean
firing   rate    although   the   network    is   inhibition-dominated
(Fig.~\ref{fig:three} {\bf a-c-d}).  The  increase in mean firing rate
with  synaptic coupling  is  therefore  a signature  of  the onset  of
fluctuating activity in this class of excitatory-inhibitory networks.

In   summary,   intrinsically   generated  fluctuating   activity   in
deterministic  excitatory-inhibitory   networks  can  be  equivalently
described by  approximating the dynamics with  a stationary stochastic
process.  Here  we  stressed  that the  mean  and  the
autocorrelation of this process are  strongly coupled and need to be
determined self-consistently.

\paragraph*{Two regimes of fluctuating activity}
The  mean field approach  revealed that,  above the  critical coupling
$J_C$,  the  network  generates  fluctuating but  stable,  stationary
activity.   The dynamical  systems  analysis, however,  showed that  the
dynamics  of an equivalent  linearized network  are  unstable and
divergent for identical parameter values.  The stability  of  the fluctuating  activity is  therefore
necessarily due to the two  non-linear constraints present in the system:
the requirement  that firing rates  are bounded from below by $0$ (i.e. positive), and  the requirement
that firing rates are limited by an upper bound $\phi_{max}$.

\begin{figure}[h!]
	\begin{adjustwidth}{-0.3in}{-0.3in} 
		\centering
		\includegraphics{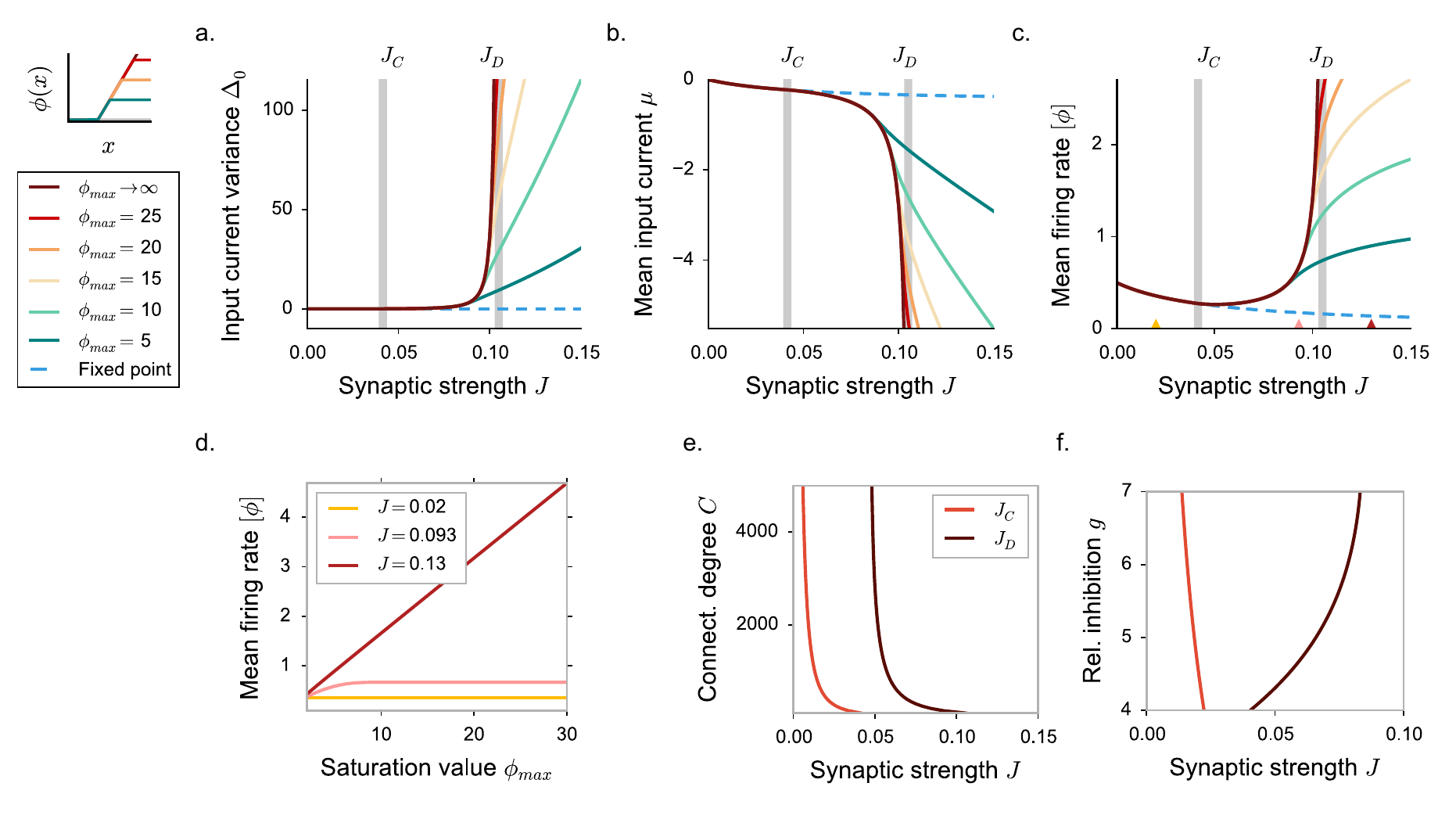}
		
		\caption{{\bf Appearance of three dynamical regimes in excitatory-inhibitory rate networks}, dynamical mean field predictions. Threshold-linear activation function saturating at different values of the upper bound $\phi_{max}$. \textbf{a-b-c.} DMF characterization of the statistics for different values of the saturation value $\phi_{max}$. In \textbf{a}, input current variance, in \textbf{b}, input current mean, in \textbf{c}, mean firing rate. Vertical grey lines indicate the critical couplings $J_C$ and $J_D$.  \textbf{d.} Mean firing rate dependence on the upper bound $\phi_{max}$, for three coupling values corresponding to the three different dynamical regimes (indicated by triangles in panel {\bf c}). Dots show time and population averages performed on 4 realizations of simulated networks, $N=6000$. Choice of the parameters: $g=5$, $C=100.$ \textbf{e-f.} Phase diagram of the dynamics:  dependence on the connectivity in-degree $C$ and on the inhibition dominance parameter $g$. All other parameters are kept fixed. }
		\label{fig:tworeg}
	\end{adjustwidth}
\end{figure}

In order to isolate the two contributions, we examined how the amplitude of
fluctuating activity depends on the  upper bound on firing rates $\phi_{max}$.
Ultimately, we take this bound to infinity, leaving the activity
unbounded.   Solving  the  corresponding  DMF equations  revealed  the
presence  of  two   qualitatively  different  regimes  of  fluctuating
activity above $J_c$ (Fig.~\ref{fig:tworeg}).

For intermediate  coupling values,  the magnitude of  fluctuations and
the  mean  firing   rate  depend  only  weakly  on   the  upper  bound
$\phi_{max}$. In particular, for  $\phi_{max} \to \infty$ the dynamics
remain stable and bounded. The  positive
feedback that  generates the linear  instability is dominantly  due to
negative, inhibitory interactions multiplying positive firing rates in
the linearized model.
In this regime, the requirement that firing
rates are  positive, combined  with dominant inhibition,  is
sufficient  to  stabilize  this feedback and the fluctuating  dynamics.  

For larger coupling values, the  dynamics depend strongly on the upper
bound  $\phi_{max}$. As  $\phi_{max}$ is  increased, the  magnitude of
fluctuations  and  the  mean  firing rate  continuously  increase  and
diverge for  $\phi_{max} \to \infty$.  For large coupling  values, the
fluctuating dynamics  are therefore stabilized by the  upper bound and
become unstable in absence of saturation, even though inhibition is globally stronger than excitation.

Fig.~\ref{fig:tworeg} {\bf d} summarizes the qualitative changes in the dependence on the upper bound $\phi_{max}$ . In the fixed point regime, mean inputs are suppressed by inhibition, and they correspond to the low-gain region of $\phi(x)$, which is independent of $\phi_{max}$. Above $J_C$, in the intermediate regime, the solution rapidly saturates to a limiting value. In the strong coupling regime, the mean firing rate, as well as the mean input $\mu$, and its standard deviation $\sqrt{\Delta_0}$ grow linearly with the upper bound $\phi_{max}$. We observe that when $\phi_{max}$ is large, numerically simulated mean activity show larger deviations from the theoretically predicted value, because of larger finite size effects (for a more detailed discussion, see \emph{Methods}).

The two regimes of fluctuating activity are characterized by different
scalings  of   the  first-   and  second-order  statistics   with  the
upper-bound  $\phi_{max}$.   In the  absence  of  upper  bound on  the
activity, i.e. in  the limit $\phi_{max} \to \infty$,  the two regimes
are  sharply separated  by  a second  ``critical''  coupling $J_D$:  below
$J_D$, the network reaches  a stable fluctuating steady-state and DMF
admits   a  solution;  above   $J_D$,  the   network  has   no  stable
steady-state, and  DMF admits no solution.  $J_D$  corresponds to the
value of the coupling for which the DMF solution diverges, and can be
determined  analytically (see  \emph{Methods}).  For  a  fixed, finite
value of  the upper  bound $\phi_{max}$, there  is however no sign of
transition  as the  coupling is  increased past  $J_D$. Indeed,  for a
fixed $\phi_{max}$,  the network  reaches a stable  fluctuating steady
state  on  both sides  of  $J_D$,  and  no qualitative  difference  is
apparent between these two  steady states. The difference appears only
when  the value  of  the  upper bound  $\phi_{max}$  is varied.  $J_D$
therefore separates  two dynamical regimes in which  the statistics of
the activity scale differently  with the upper-bound $\phi_{max}$, but
for  a  fixed,  finite  $\phi_{max}$  it does  not  correspond  to  an
instability. The second ``critical''  coupling $J_D$ is therefore qualitatively different from the critical coupling $J_c$, which is associated with an instability for any value of $\phi_{max}$.

The value of $J_D$  depends both  on the
relative strength of inhibition $g$,  and the total number of incoming
connections  $C$.  Increasing  either  $g$ or  $C$
increases  the  total variance  of  the  interaction matrix  $J_{ij}$,
shifting the instability of the homogeneous fixed point to lower couplings.  The
size of  the intermediate fluctuating  regime however  depends
only weakly  on the  number  of incoming  connections  $C$ (Fig.~\ref{fig:tworeg} {\bf e}). 
In  contrast,
increasing  the   relative  strength  of   inhibition  diminishes  the
influence  of the  upper bound  and  enlarges the  phase space  region
corresponding  to  the  intermediate  regime, where  fluctuations  are
stabilized intrinsically by  recurrent inhibition (Fig.~\ref{fig:tworeg} {\bf f}). The second critical
coupling  $J_D$ is  in particular  expected to  increase with  $g$ and
diverge  for purely  inhibitory  networks.  However, for  very large relative
inhibition, numerical simulations show strong deviations from
DMF predictions,  due to the  breakdown of the  Gaussian approximation
which  overestimates  positive feedback (see \emph{Methods}).

In summary,  the two non-linearities  induced by the  two requirements
that the firing rates are positive and bounded play asymmetrical roles
in stabilizing fluctuating dynamics. In excitatory-inhibitory networks
considered here,  this asymmetry leads to  two qualitatively different
fluctuating regimes.

\paragraph*{The effect of spiking noise}
We  next   investigated  whether  the two different fluctuating regimes described above can be still observed
when spiking noise is  added to the dynamics.  Following \cite{Kadmon,Goedeke2016},
we  added  a  Poisson  spiking  mechanism  on  the  rate  dynamics  in
Eq.~\eqref{eq:dyn}, and let the  different units interact through spikes
(see \emph{Methods}).  Within  a mean field  approach, interaction
through spikes  lead to an additive  white noise term  in the dynamics  \cite{Kadmon,Goedeke2016}. To determine  the effect of  this additional
term  on the  dynamics,  we first  treated  it as  external noise  and
systematically varied its amplitude as a free parameter.

\begin{figure}[h!]
	\begin{adjustwidth}{-0.3in}{-0.3in} 
		\centering
		\includegraphics{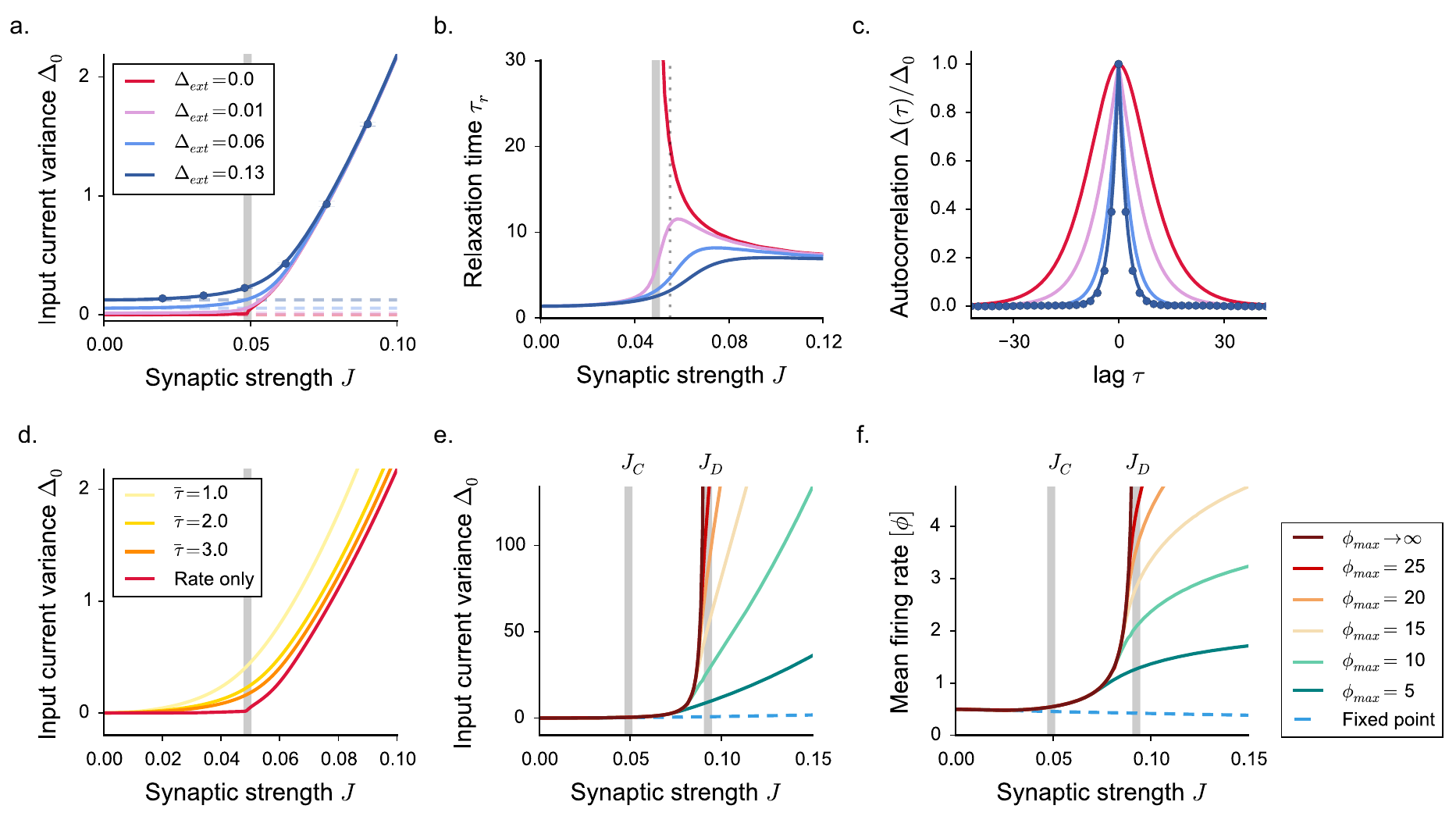}
		\caption{ {\bf Statistical description of  the activity in excitatory-inhibitory networks with external and spiking noise.} The dynamical mean field results are shown in full lines, numerical simulations as points. \textbf{a.} Input current variance in presence of external noise, for increasing values of the noise amplitude (white noise, variance equal to $2\Delta_{ext}$). Blue dots: results of numerical simulations for $\Delta_{ext}=0.13$, $N=7500$, average of 4 realizations of the synaptic matrix. The grey vertical line shows the critical coupling $J_C$ in the deterministic model. Dashed lines indicate the statistics of an effective fixed point, where the only variance is generated by the noise contribution $\Delta_{ext}$. The fixed point firing rate is computed as a Gaussian average, with the mean given by the fixed point $x_0$ and the variance provided solely by the noise term. The deflection from the effective fixed point underlines an internal amplification of noise produced by network feedback. \textbf{b.} Fluctuations relaxation time, measured as the auto-correlation $\Delta(\tau)$ full width at half maximum. \textbf{c.} Normalized auto-correlation for fixed $J$ and different levels of noise.  The corresponding coupling value is indicated by the dotted vertical gray line in panel \textbf{b}. \textbf{d.} Input variance in a network with spiking dynamics, where spikes are generated according to inhomogeneous Poisson processes. Increasing the time constant of  rate dynamics $\bar{\tau}$ (see Eq.~\eqref{eq:rate-spk} in \emph{Methods}) decreases the amplitude of spiking noise. \textbf{e-f.}  Appearance of the three dynamical regimes in a network with spiking noise: input current variance and mean firing rate for different saturation values $\phi_{max}$. Choice of the parameters: $g=4.1$, $C=100$.}
		\label{fig:noise}
	\end{adjustwidth}
\end{figure}

The main effect of noise is to induce fluctuations in the activity for
all values of network parameters (Fig.~\ref{fig:noise} {\bf a}).   As a result, in presence of noise,
the  sharp transition  between  constant and  fluctuating activity  is
clearly  lost  as previously shown \cite{Kadmon,Goedeke2016}.   The  feedback  mechanism   that  generates  intrinsic
fluctuations nevertheless still operates and  strongly amplifies the
fluctuations induced by external noise.

The DMF framework can be extended to include external
noise and  determine the additional variability  generated by network
feedback (\cite{Kadmon,Goedeke2016}, see also \emph{Methods}). When the coupling $J$ is small, the temporal fluctuations in
the activity  are essentially generated  by the filtering  of external
noise.  Beyond the  original transition  at $J_C$, instead, 
when  the feedback fluctuations grow rapidly with  synaptic coupling, 
the contribution of external  noise becomes rapidly
negligible with respect to the intrinsically-generated fluctuations (Fig.~\ref{fig:noise} {\bf a}).

As shown in earlier studies \cite{Kadmon,Goedeke2016}, a dramatic effect of introducing  external noise is a strong reduction
of    the     timescale    of    fluctuations     close    to    $J_C$. In  absence of  noise, just above  the fixed
point  instability at  $J_C$, purely  deterministic rate  networks are
characterized by the onset of infinitely slow fluctuations. These slow
fluctuations are however of  vanishingly small magnitude, and strongly
sensitive  to external  noise.  Any finite  amount  of external  noise
eliminates  the  diverging  timescale.   For weak  external  noise,  a
maximum in the  timescale can be still seen close  to $J_C$, but it
quickly disappears as the magnitude  of noise is increased. For modest
amounts of  external noise, the timescale of  the fluctuating dynamics
becomes     a    monotonic     function    of     synaptic    coupling
(Fig.~\ref{fig:noise} {\bf b}).

While in presence of external  noise there is therefore no formal critical phase
transition,    the   dynamics    still   smoothly    change  from
externally-generated   fluctuations   around    a   fixed   point   into
intrinsically-generated,  non-linear  fluctuations.   This  change  of
regime is not necessarily reflected  in the timescale of the dynamics,
but  can clearly  be seen  in  the excess  variance, and  also in  the
first-order  statistics  such as  the  mean-firing  rate, which  again
strongly increases with coupling. Moreover,  we found that the existence of the second
fluctuating regime  is totally insensitive to noise:  above the second
critical coupling $J_D$, the activity  is only stabilized by the upper
bound on the firing rates, and  diverges in its absence. In that parameter
region, intrinsically-generated fluctuations diverge, and the external
noise contributes only a negligible amount.

We  considered  so  far the  effect  of  an  external white  noise  of
arbitrary  amplitude. If that  noise represents  spiking interactions,
its variance  is however  not a free  parameter, but instead  given by
$J^2(C_E+g^2C_I)[\phi]/\bar{\tau}$.  In particular,  the  amplitude of
spiking noise increases  both with the synaptic coupling  and with the
mean firing rate $[\phi]$, which  itself depends on the coupling  and fluctuations as pointed out above. As a
result, the  amplitude of the spiking noise  dramatically increases in
the  fluctuating  regime  (Fig.~\ref{fig:noise}  {\bf  d}).  When  $J$
becomes close to the second critical coupling $J_D$, the spiking noise
however still  contributes only weakly  to the total variance  (see in
\emph{Methods}),  and  the  value  of  $J_D$ is  not  affected  by  it
(Fig.~\ref{fig:noise} {\bf e}). The amplitude of spiking noise is also
inversely  proportional the  timescale $\bar{\tau}$  of  the dynamics
(see Eq.~\eqref{eq:rate-spk} in \emph{Methods}). Slower dynamics  tend
to smooth out fluctuations due to spiking inputs (Fig.~\ref{fig:noise}
{\bf d}), reduce the amount of spiking and noise and therefore favor the appearance of slow fluctuations close to the critical coupling $J_c$ \cite{Kadmon}.

In conclusion, the main   findings reported above, the influence of
intrinsically  generated  fluctuations   on  mean  firing rate,  and  the
existence of two different  fluctuating regimes are still observed in
presence of  external or  spike-generated noise. In  particular, above
the  second  transition, intrinsically  generated  fluctuations can  be
arbitrarily strong  and therefore play the dominant  role with respect
to external or spiking noise.

\paragraph*{Purely inhibitory networks}

\begin{figure}[h!]
	\begin{adjustwidth}{-0.3in}{-0.3in} 
		\centering
		\includegraphics{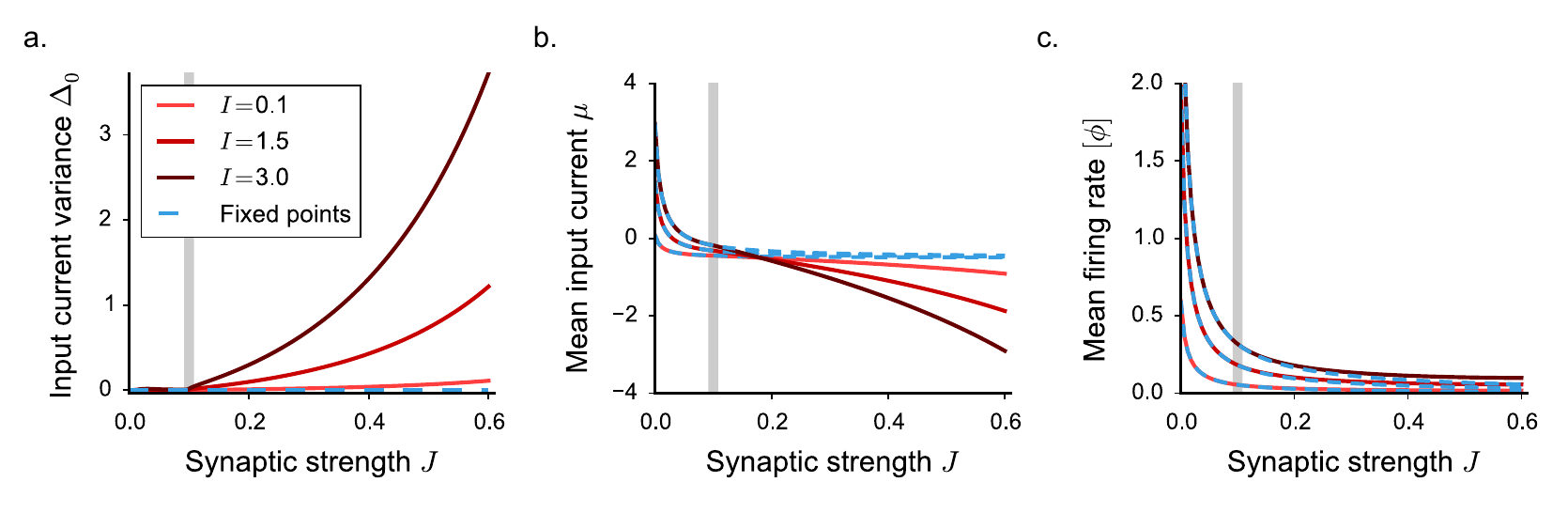}
		\caption{
			{\bf Statistical description of the activity in purely inhibitory networks.} Results of the dynamical mean field theory (obtained through setting $C_E=0$ and $g=1$) for different values of the excitatory external current $I$.
			\textbf{a.} Input current variance, \textbf{b.} mean current and \textbf{c.} mean firing rate as a function of the synaptic coupling $J$. Vertical grey lines indicate the critical value $J_C$. }
		\label{fig:inh}
	\end{adjustwidth}
\end{figure}

To identify  the  specific role  of  excitation in  the
dynamics  described above,  we  briefly consider  here  the case  of
networks consisting  of a  single inhibitory population.   Purely
inhibitory networks display a transition  from a fixed point regime to
chaotic   fluctuations   \cite{Kadmon,Harish}.    The   amplitude   of
fluctuations  appears   to  be  in   general  much  smaller   than  in
excitatory-inhibitory  networks,  but   increases  with  the  constant
external  current $I$  (Fig.~\ref{fig:inh} {\bf  a}).  In contrast  to our findings  for networks in which  both excitation
and   inhibition   are   present,   in  purely   inhibitory   networks
intrinsically generated  fluctuations lead to a very  weak increase in
mean firing rates compared to the fixed point (Fig.~\ref{fig:inh} {\bf
	b-c}).  This effect  can be  understood  by noting  that within  the
dynamical   mean field   theory,   the   mean   rate   is   given   by
$(\mu-I)/J(C_E-gC_I)$  (Eq.~\eqref{eq:dmf-mu}). The  term  $C_E-gC_I$ in
the denominator determines the sensitivity  of the mean firing rate to
changes  in  mean  input. This  term  is  always  negative as  we  are
considering inhibition-dominated  networks, but its  absolute value is
much  smaller in  presence  of excitation,  i.e.  when excitation  and
inhibition  approximately  balance,   compared  to  purely  inhibitory
networks.  As  the   onset  of  intrinsically  generated  fluctuations
modifies the value of the mean  input with respect to its value in the
fixed point  solution (Fig.~\ref{fig:three} {\bf c}, Fig.~\ref{fig:inh} {\bf b}), this simplified argument explains  why the mean
firing  rates in  the inhibitory  network are  much less  sensitive to
fluctuations than in the excitatory-inhibitory case.

Moreover the second fluctuating regime  found in EI networks does not
appear  in  purely inhibitory  networks.   Indeed,  the divergence  of
first-  and  second-order  statistics  that  occurs  in  EI  networks
requires  positive  feedback  that  is  absent  in  purely  inhibitory
networks.  Note that for purely inhibitory, sparse networks, important
deviations  can exist at  very large  couplings between  the dynamical
mean  field theory  and  simulations (see  \emph{Methods}  for a  more
detailed discussion).

The two  main findings reported above, the strong influence of intrinsically
generated  fluctuations on  mean firing rate,  and the  existence  of two
different  fluctuating  regimes   therefore  critically  rely  on  the
presence of excitation in the network.

\subsection*{Extensions to more general classes of networks}

\paragraph*{General excitatory-inhibitory (EI) networks}
In  the class  of  networks  we investigated  so  far, excitatory  and
inhibitory units received  statistically equivalent inputs. Under this
assumption, the  network dynamics are  characterized by a  single mean
and  variance for  both excitatory  and inhibitory  populations, which
considerably simplifies the mean field description. Here we relax this
assumption  and show  that the  properties   of intrinsically
generated fluctuations described so far do not critically depend on it.

We consider a more general  class of networks,  in which synaptic
connections are arranged in a block matrix:
\begin{equation}
	\mathrm{J}=J\left(
	\begin{array}{c|c}
		\mathrm{J_{EE}}  &  \mathrm{J_{EI}} \\ \hline
		\mathrm{J_{IE}} &  \mathrm{J_{II}}
	\end{array}\right)
\end{equation}
where each block $\mathrm{J_{kk'}}$  is a sparse matrix, containing on
each row $C_{kk'}$ non-zero  entries of value $j_{kk'}$. The parameter
$J$  represents a  global scaling  on  the intensity  of the  synaptic
strength. For  the sake  of simplicity, we  restrict ourselves  to the
following  configuration: each  row of  $\mathrm{J}$  contains exactly
$C_E$  non-zero excitatory  entries in  the blocks  of  the excitatory
column,  and  exactly  $C_I$  inhibitory  entries  in  the  inhibitory
blocks. Non-zero elements are  equal to $j_E$ in $\mathrm{J_{EE}}$, to
$-g_Ej_E$ in $\mathrm{J_{EI}}$, to  $j_I$ in $\mathrm{J_{IE}}$, and to
$-g_Ij_I$  in $\mathrm{J_{II}}$.  The  previous case  is recovered  by
setting $j_E=j_I=1$ and $g_E=g_I$.

The network admits a fixed point in which the activities are different for excitatory and inhibitory units, but homogeneous within the two populations. This fixed point is given by:

\begin{equation}
	\left(
	\begin{array}{c}
		x_0^E  \\
		x_0^I 
	\end{array}\right)=J
	\left(
	\begin{array}{c}
		j_E(C_E \phi(x_0^E)-g_E C_I \phi(x_0^I))  \\
		j_I(C_E \phi(x_0^E)-g_I C_I \phi(x_0^I))
	\end{array}\right) \label{eq:EIgen-fp}	
\end{equation}
where $x_0^E$ and $x_0^I$ are the fixed-point inputs to the two populations.

The linear stability of the fixed point is determined by the eigenvalues of the matrix:
\begin{equation}
	\mathrm{S}=J\left(
	\begin{array}{c|c}
		\phi'(x_0^E)\mathrm{J_{EE}}  &  \phi'(x_0^I)\mathrm{J_{EI}} \\ \hline
		\phi'(x_0^E)\mathrm{J_{IE}} &  \phi'(x_0^I)\mathrm{J_{II}}.
	\end{array}\right)\label{eq:EIgen-stab}
\end{equation}
The fixed point is stable if  the real part of all the eigenvalues is smaller than one. As for simple, column-like EI matrices, the eigenspectrum of $\mathrm{S}$ is composed of a discrete and a densely distributed part, in which the bulk of the eigenvalues are distributed on a circle in the complex plane \cite{Aljadeff1, Aljadeff2, AljadeffEig}. The discrete component consists instead of two eigenvalues, which in general can be complex, potentially inducing various kinds of fixed point instabilities (for the details, see \emph{Methods}).
As in the previous paragraphs, we consider a regime where both $g_E$ and $g_I$ are strong enough to dominate excitation, and the outlier eigenvalues have negative real part. In those conditions, the first instability to occur is the chaotic one, where the radius of the complex circle of the eigenspectrum crosses unity.
This radius increases with the overall coupling $J$, defining a critical value $J_C$ where the fixed point loses stability.

\begin{figure}[h!]
	\begin{adjustwidth}{-0.3in}{-0.3in} 
		\centering
		\includegraphics{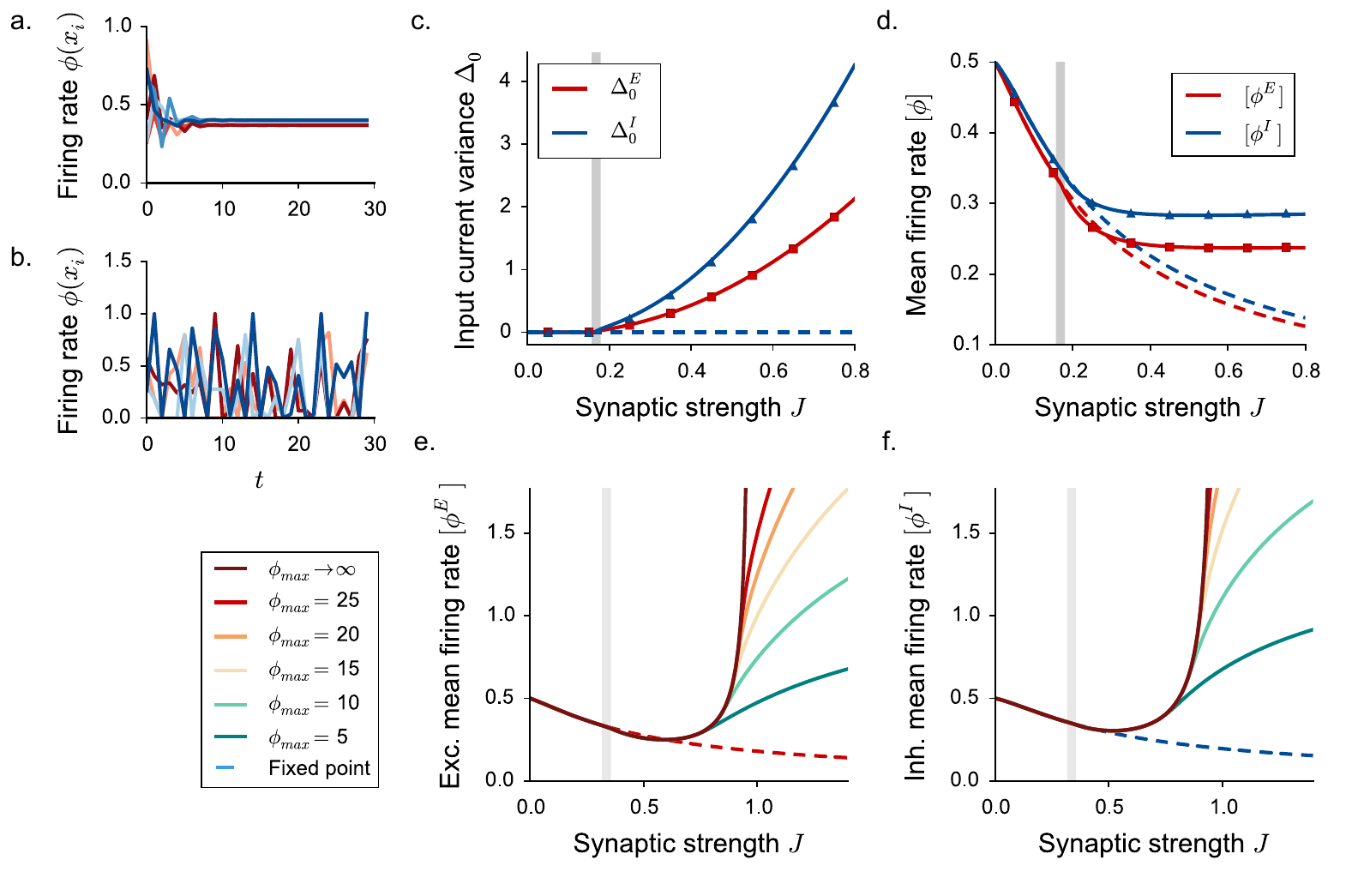}
		\caption{{\bf Fluctuating dynamics in more general networks where excitatory and inhibitory neurons are not statistically equivalent.} Discrete-time rate evolution. \textbf{a-b.} Network discrete-time activity: numerical integration of the Eq.~\eqref{eq:discrete}, firing rates of randomly selected units. Excitatory neurons are plotted in the red scale, inhibitory ones in the blue one. $N=1000$. In \textbf{a}, $J<J_C$; in \textbf{b}, $J>J_C$. \textbf{c-d.} Statistical characterization of network activity, respectively in terms of the input variance and the mean firing rate. Dynamical mean field results are shown in full lines. Dashed lines: fixed points. Dots: numerical simulations, $N=7500$, average over 3 realizations. Vertical grey lines indicate the critical value $J_C$. $\phi_{max}=1$. \textbf{e-f.} Mean firing rate for different values of the saturation $\phi_{max}$, in the excitatory and the inhibitory population. Choice of the parameters: $j_E=0.1$, $j_I=1.5 j_E$, $g_E=4.5$, $g_I=4.2$, $C=100$.}
		\label{fig:2pop}
	\end{adjustwidth}
\end{figure}

Dynamical mean field equations for the fluctuating regime above the
instability are,  in this general case,  much harder to  solve as they
now involve two means and two auto-correlation functions, one for each populations \cite{Kadmon,Harish}.
For that reason, we restrict ourselves to a slightly different dynamical system with discrete-time  evolution:
\begin{equation}
	x_i(t+1)=\sum_{j=1}^{N} J_{ij}\phi(x_j(t)).
	\label{eq:discrete}
\end{equation}
Such a network corresponds to  extremely fast dynamics with no current
filtering (Fig.~\ref{fig:2pop} {\bf a-b}).      Previous    works     \cite{Molgedey, Cessac, Cessac2, Toyoizumi} have studied that class of models in case of synaptic
matrices that lacked EI separation, and for activation functions that
were symmetric. These works pointed out strong analogies with the dynamics emerging in continuous time \cite{Sompolinsky}. Discrete-time dynamics can however induce a new, period-doubling bifurcation when inhibition is strong. We therefore restrict the analysis to  a regime where inhibition is dominating but not excessively strong.
Notice that in general, outside the range of parameters considered in this analysis, we expect generic EI networks to display a richer variety of dynamical regimes.

To  begin with,  we observe  that  the fixed-point (Eq.~\eqref{eq:EIgen-fp})  and its  stability
conditions (Eq.~\eqref{eq:EIgen-stab})  are identical  for continuous  and discrete  dynamics. For
discrete time,  the DMF  equations are however  much simpler  than for
continuous  dynamics,  and  can  be  easily fully  solved  even  if  the  two
populations  are characterized  now by  different values  of  mean and
variance.

Solving the DMF equations  confirms that the transition to chaos in this class of models is characterized by the same qualitative features as before (Fig. \ref{fig:2pop} {\bf c-d}). 
As the order parameter $J$ is increased, the means and the variances of both the E and the I population display a transition from the fixed point solution to a fluctuating regime characterized by positive variance $\Delta_0$ and increasing mean firing rate.
By smoothly increasing the upper bound of the saturation function $\phi_{max}$ as before, we find a second critical value $J_D$ at which the firing activity of both populations diverge (Fig.~\ref{fig:2pop} {\bf e-f}). We conclude that the distinction in three regimes reported so far can be extended to discrete-time dynamics; in this simplified framework, our results extend to more general EI connectivity matrices.

\paragraph*{Connectivity with stochastic in-degree}
We now turn to networks in which the number of incoming connections is
not fixed for all the  neurons, but fluctuates stochastically around a
mean  value $C$.   We consider  a  connectivity scheme  in which  each
excitatory (resp.  inhibitory) neuron  makes a connection  of strength
$J$ (resp. $-gJ$) with probability $C/N$.

In  this class  of networks,  the number  of incoming  connections per
neuron has  a variance  equal to  the mean. As  a consequence,  in the
stationary  state, the  total input  strongly varies  among  units. In
contrast to the case of a  fixed in-degree, the network does not admit
an homogeneous, but a heterogeneous fixed point in which different units reach different equilibrium values depending on the details of the connectivity. 

The  dynamical mean  field approach  can  be extended  to include  the
heterogeneity generated by the variable number of incoming connections
\cite{Toyoizumi,  Kadmon, Harish}. As  derived in  \emph{Methods}, the
stationary  distributions are  now described  by a  mean and  a static
variance  $\Delta_0$  that   quantifies  the  static,  quenched  noise
generated  by variations in  the total  input among  the units  in the
population. These two quantities obey:
\begin{equation}
	\begin{split}
		&\mu= J(C_E-gC_I)[\phi]+I,\\
		&\Delta_0= J^2(C_E+g^2C_I)[\phi^2].
	\end{split}
\end{equation}

\begin{figure}[h!]
	\begin{adjustwidth}{-0.3in}{-0.3in} 
		\centering 
		\includegraphics{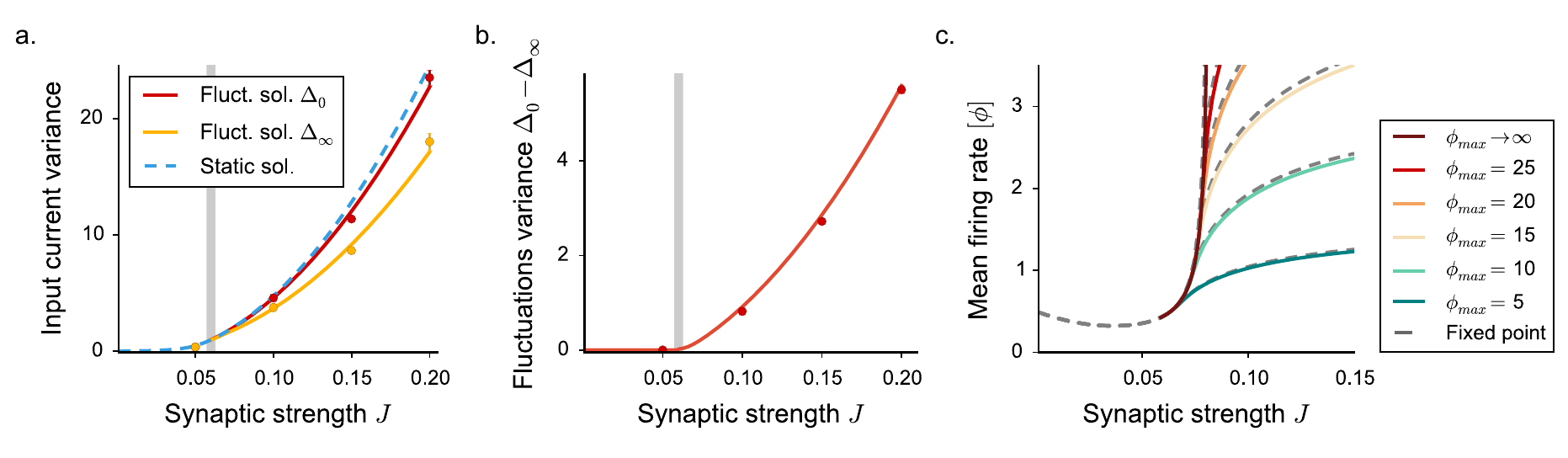}
		\caption{{\bf Mean field characterization of the activity in networks with stochastic in-degree.} The dynamical mean field results are shown in full lines, numerical simulations as points. \textbf{(a)} Total input current variance $\Delta_0$. The heterogeneity in the connectivity induces an additional quenched variance $\Delta_{\infty}$ (shown in dashed blue for the fixed point, and yellow for the fluctuating solution, where it corresponds to $\Delta_0$). Red (resp. yellow) points show time and population averages of $\Delta_0$ (resp. $\Delta_{\infty}$) performed on 3 realizations of simulated networks, $N=6500$. \textbf{(b)} Isolated contribution of temporal fluctuations to the variance. \textbf{(c)} Mean firing rate, for different values of the saturation $\phi_{max}$. Grey dashed lines indicate the stationary solution, becoming a thick colored line, corresponding to the chaotic phase, at $J_C$. Choice of the parameters: $g=5$, $C=100$, $\phi_{max}=2$.}
		\label{fig:randomc}
	\end{adjustwidth}
\end{figure}

The  stationary   solution  loses   stability  at  a   critical  value
$J=J_C$. In the  strong coupling regimes, DMF predicts  the onset of a
time-dependent solution with a decaying autocorrelation function, with
initial condition  $\Delta_0$ and asymptotic  value $\Delta_{\infty}$.
The values  of $\mu$, $\Delta_0$ and  $\Delta_{\infty}$ are determined
as    solution    of    a    system   of    three    equations    (see
Eqs.~\eqref{eq:randomc1_mem},          \eqref{eq:randomc2_mem}         and
\eqref{eq:randomc3_mem}  in   \emph{Methods}).   In  this   regime,  the
effective  amplitude   of  temporal  fluctuations  is   given  by  the
difference   $\Delta_0-\Delta_{\infty}$  (Fig.~\ref{fig:randomc}  {\bf
	b}).    A   non-zero  value   of   $\Delta_{\infty}$  reflects the variance of mean activity across the population:   the
the activity of different units fluctuates around different mean values because of the heterogeneity in the connectivity.
Note moreover  that
because    the    static     variance    increases strongly   with    coupling
(Fig.~\ref{fig:randomc}  {\bf a}),  the mean  activity for  the static solution  increases with
coupling , in contrast to the fixed in-degree case.  In the fluctuating regime,  as the
additional  temporal  variance  $\Delta_0-\Delta_{\infty}$ is  weaker
than the  static variance $\Delta_{\infty}$,  temporal fluctuations do
not lead to  an increase in mean firing  rate with respect to
the static solution (Fig.~\ref{fig:randomc} {\bf c}), in contrast to our findings for the fixed in-degree case.

Fig.~\ref{fig:randomc} {\bf  c} displays  the dependence on  the upper
bound $\phi_{max}$ of  the mean field solution.  Above  $J_C$, an  intermediate regime exists  where the
activity  is stabilized  by  inhibition, and  remains  finite even  in
absence of upper bound. For couplings above a second critical coupling
$J_D$,  the   dynamics  are  stabilized   only  by  the   upper  bound
$\phi_{max}$. Networks with variable in-degree therefore show the same three dynamical regimes as networks with fixed degree.

\subsection*{Comparing rate and integrate-and-fire networks}

For excitatory-inhibitory  networks of threshold-linear  rate units, we
have identified two different regimes of fluctuating activity. In this
section, we  show that networks of spiking,  leaky integrate-and-fire (LIF)
neurons display the signatures characteristic of these two regimes. To
link threshold-linear rate networks to LIF networks, we first consider
a modified rate model directly related to LIF networks \cite{Ostojic}, and then perform simulations of spiking LIF networks.

\paragraph*{Rate networks with an LIF transfer function}
We focus  again on  the fixed in-degree  synaptic matrix in  which the
inputs  to   excitatory  and  inhibitory   neurons  are  statistically
equivalent, but consider a rate network in which the dynamics are now given by:

\begin{equation}
	\dot{\phi_i}(t)=-\phi_i(t)+F(\mu_i(t), \sigma_i(t)) \label{eq:rate-lif1}
\end{equation}
where:
\begin{equation}
	\begin{split}
		&\mu_i(t)=\mu_0+\tau_m \sum_j J_{ij} \phi_j(t)\\
		&\sigma^2_i(t)=\tau_m \sum_j J_{ij}^2 \phi_j(t).
	\end{split}
\end{equation}
Here $\phi_i$  is the firing rate  of unit $i$, $\mu_0$  is a constant
external input, and $\tau_m=20$ ms  is the membrane time constant. The
function  $F(\mu,\sigma)$  is the  input-output  function  of a  leaky
integrate-and-fire neuron receiving a  white-noise input of mean $\mu$
and variance $\sigma$ \cite{Siegert}:
\begin{equation}
	F(\mu, \sigma^2)=\left[\tau_{rp}+2 \tau_m \int_{\frac{V_r-\mu}{\sigma}} ^{\frac{V_{th}-\mu}{\sigma}} \diff u e^{u^2} \int_{-\infty}^{u} \diff \nu e^{-\nu^2}\right]^{-1}
\end{equation}
where $V_{th}$ and $V_r$ are the threshold and reset potentials of the
LIF neurons, and $\tau_{rp}$ is the refractory period.

The  firing-rate model defined  in Eq.~\eqref{eq:rate-lif1}  is directly
related  to  the  mean  field  theory  for  networks  of  LIF  neurons
interacting through instantaneous synapses \cite{Brunel99, Brunel2000,
	Ostojic}.   More  specifically,  the  fixed point  of  the  dynamics
defined  in  Eq.~\eqref{eq:rate-lif1} is  identical  to the  equilibrium
firing rate  in the classical asynchronous  state of a  network of LIF
neurons   with   an  identical   connectivity   as   the  rate   model
\cite{Brunel99, Brunel2000}.  Eq.~\eqref{eq:rate-lif1}  can then be seen
as    simplified    dynamics    around    this    equilibrium    point
\cite{OstojicBrunel2011,Shaffer2013}.  A linear stability  analysis of
the fixed point  for the rate model predicts  an instability analogous
to the one found in  threshold-linear rate models.  A comparison with a
network of LIF  neurons shows that this instability  predicts a change
in the  dynamics in the corresponding spiking  network, although there
may  be  quantitative  deviations  in  the  precise  location  of  the
instability \cite{Ostojic,Engelken,Ostojic2015}.

The   dynamics  of   Eq.~\eqref{eq:rate-lif1}  have   been  analytically
investigated   only  up   to  the   instability   \cite{Ostojic}.   To
investigate   the    dynamics   above   the    instability,   we   set
$x_i(t)=\sum_{j=1}^{N}J_{ij}\phi_j(t)$,  and rewrite  the  dynamics in
the more familiar form:
\begin{equation}
	\dot{x_i}(t)=-x_i(t)+\sum_{j=1}^{N}J_{ij}F(\tau_m x_j(t), \sigma_j(t)). \label{eq:rate-lif2}
\end{equation}
The main  novelty with  respect to previously  studied rate  models is
that the  input-output transfer function  $F$ depends on  the standard
deviation  $\sigma_j$  of  the  input  current  to  the  unit  $j$.  A
dependence  on  a  time-varying  $\sigma_j$ is  however  difficult  to
include in the dynamical mean field approach. As a step forward, we  fix $\sigma_j$ to its average value independent
of  $j$ and  time, which  corresponds to  substituting all  the firing
rates with a constant effective value $\bar{\phi}$:
\begin{equation}
	\sigma^2\sim \tau_m\sum_j J_{ij}^2\bar{\phi}=\tau_mJ^2(C_E+g^2 C_I)\bar{\phi}.
\end{equation}
With this substitution, we are back  to a classical rate model with an
LIF transfer function.  Quantitatively  the dynamics of that model are
not identical to the model defined in Eq.~\eqref{eq:rate-lif1}, but they
can be studied using dynamical  mean field theory.  We therefore focus
on  qualitative  features of  the  dynamics  rather than  quantitative
comparisons between models.

\begin{figure}[h!]
	\begin{adjustwidth}{-0.3in}{-0.3in} 
		\centering
		\includegraphics{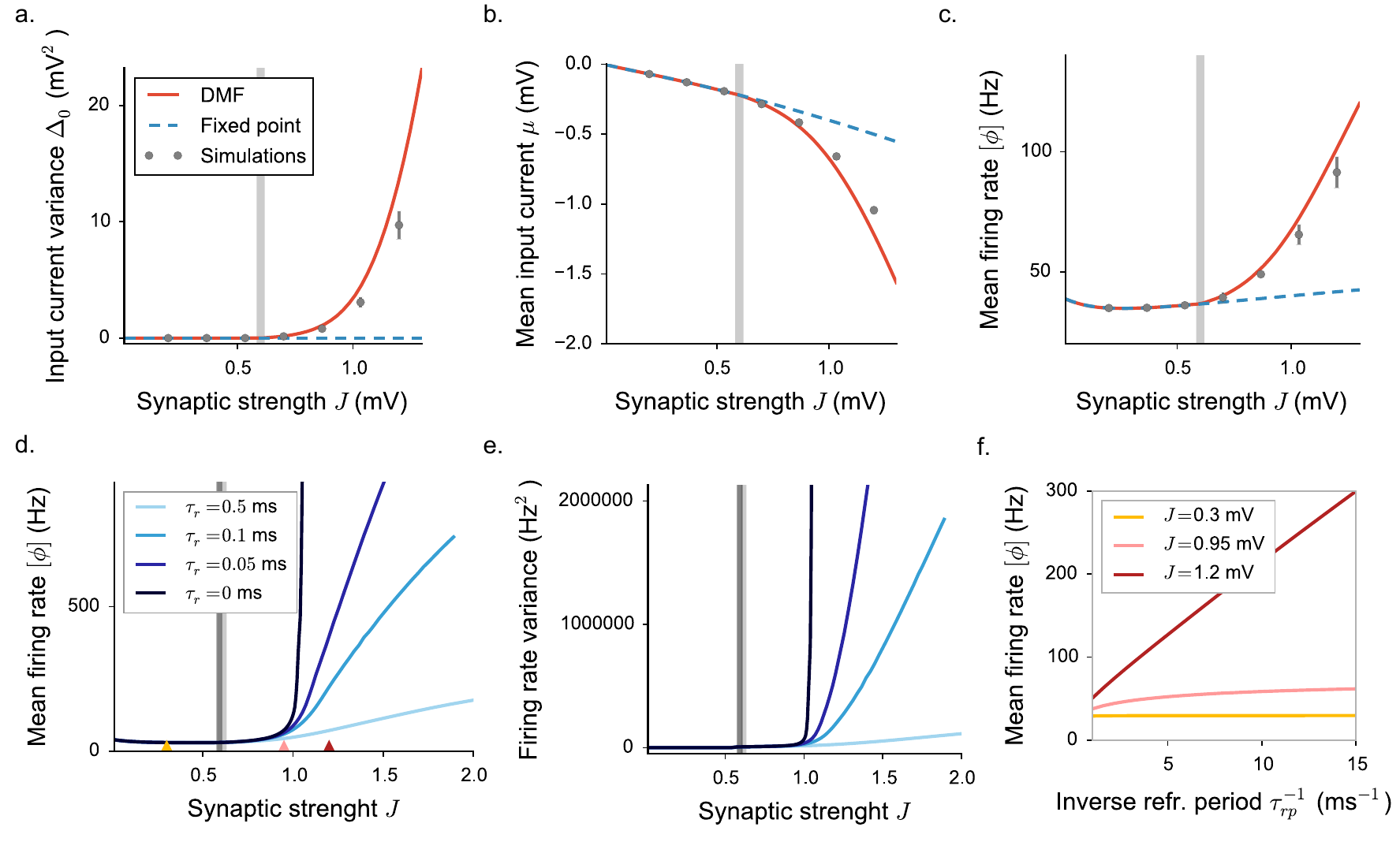}
		\caption{
			{\bf Dynamical mean field characterization of rate network activity with a LIF activation function}, where we set $\sigma^2=\tau_m J^2(C_E+g^2 C_I)\bar{\phi}$, $\bar{\phi}=20$ Hz.
			{\bf a-b-c.} Statistical characterization for $\tau_r=0.5$ ms: input variance, mean input current and mean firing rate. Grey vertical lines indicate the position of the critical coupling. Choice of the parameters: $g=5$, $C=100$. {\bf d-e.} Mean firing rate and rate standard deviation for different values of the refractory period, determining slightly different positions of the transition (grey lines). Choice of the parameters: $g=5$, $C=100$, $\mu_0=24$ mV. {\bf f.} Mean firing rate dependence on the refractory period, the inverse of which determines the saturation value of the transfer function. The three values of the synaptic coupling, indicated by triangles in {\bf c}, correspond to the three different regimes.}
		\label{fig:spk-rate}
	\end{adjustwidth}
\end{figure}

Solving the dynamical mean field  equations shows that the dynamics in
the  rate  model with  and  LIF  transfer  function are  qualitatively
similar  to the  threshold-linear rate  model studied  above.   As the
coupling strength $J$  is increased above a critical  value, the fixed
point  loses  stability,  and   a  fluctuating  regime  emerges.   The
amplitude    of    the    fluctuations   increases    with    coupling
(Fig.~\ref{fig:spk-rate} {\bf a}), and induces an increase of the mean
firing  rate with  respect to  values  predicted for  the fixed  point
(Fig.~\ref{fig:spk-rate} {\bf c}).

In the  LIF transfer function, the  upper bound on the  firing rate is
given  by the  inverse of  the  refractory period.  For that  transfer
function,  changing the  refractory period  does not  modify  only the
upper  bound, but  instead affects  the full  function.  For different
values of the refractory periods,  the fixed point firing rate and the
location of  the instability therefore  change, but these  effects are
very small for refractory periods below one millisecond.

Varying  the  refractory  period  reveals  two  different  fluctuating
regimes       as       found       in      threshold-linear       rate
models (Fig.~\ref{fig:spk-rate}   {\bf    d-e-f}).    At   intermediate
couplings, the  fluctuating dynamics  depend weakly on  the refractory
period and remain bounded if the  refractory period is set to zero. At
strong couplings, the fluctuating  dynamics are stabilized only by the
presence of the  upper bound, and diverge if  the refractory period is
set to zero.   The main difference with the  threshold-linear model is
that the additional dependence on the coupling $J$ induced by $\sigma$
on  the  transfer function  reduces  the  extent  of the  intermediate
regime.

\begin{figure}[h!]
	\begin{adjustwidth}{-0.3in}{-0.3in} 
		\centering
		\includegraphics{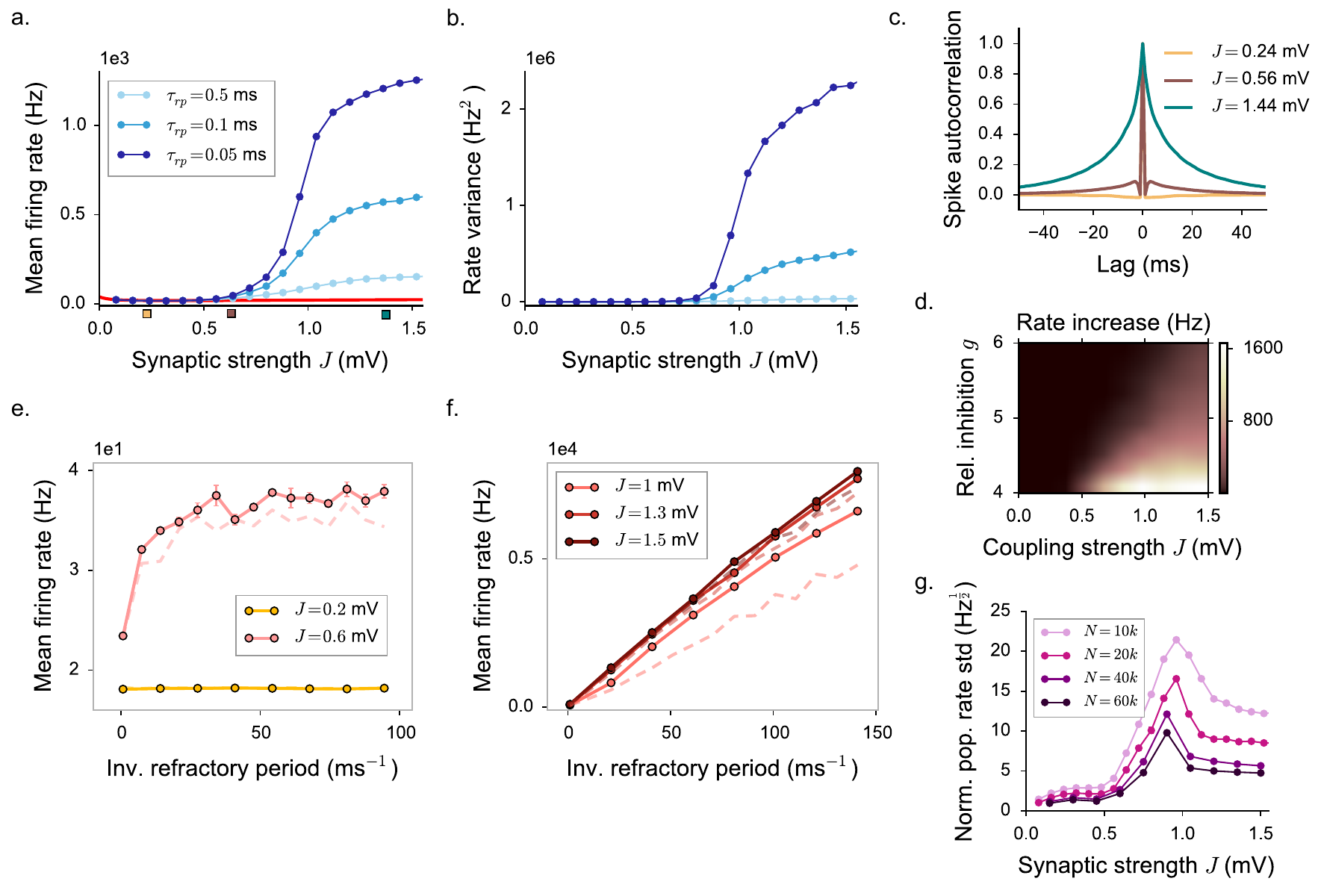}
		\caption{ {\bf Statistical characterization of  activity in a network of leaky integrate-and-fire  neurons.}  \textbf{a.} Mean firing rate. Numerical simulations ($N=20000$) are in good agreement with the LIF mean field prediction (red line) for low coupling values ($J<0.5$). For high values of $J$ ($J>0.8$),  mean firing rates diverge and becomes highly dependent on the refractory period.
			\textbf{b.} Firing rate variance, computed on instantaneous firing rates evaluated with a $50$ ms Gaussian filter. 
			\textbf{c.} Spike autocorrelation function, computed with 1 ms time bins, for three different values of the coupling $J$ ($\tau_{rp}=0.5$).
			\textbf{d.} Increase in the mean firing rate as the refractory period is decreased from 0.5 to 0.1 ms, as a function of the synaptic coupling $J$ and the inhibition strength $g$. As in the rate networks, the mean firing rate and its increase depend on the value of $g$. 
			\textbf{e-f.} Direct dependence between the mean firing rate and refractory period. Panel \textbf{e} shows the low and intermediate coupling regime. Panel \textbf{f} shows the high coupling regime.  Colored dots: simulated networks with $N=20000$. Lighter dashed lines (when visible) show the result for $N=10000$.
			\textbf{g.} Dependence on $J$ and $N$ of correlations and synchrony, quantified by the std of the population-averaged spiking rate, normalized by the square root of the mean firing rate ($\tau_{rp}=0.05$). Std is computed within a time bin of 1 ms.  
			In all the panels, choice of the parameters: $g=5$, $C=500$, $\Delta$=1.1 ms,  $\mu_0=24$ mV.	}
		\label{fig:spk}
	\end{adjustwidth}
\end{figure}

\paragraph*{Spiking networks of leaky integrate-and-fire neurons}
Having  established   the  existence  of  two   different  regimes  of
fluctuating activity in rate  networks with an LIF transfer function,
we next  consider spiking networks  of LIF neurons.  To  compare the
different regimes of activity in  spiking networks with the regimes we
found in rate networks, we performed direct numerical simulations of a
spiking LIF network.   We examined the effects of the coupling strength
and   refractory  period   on  first-   and   second-order  statistics
(Fig.~\ref{fig:spk}  {\bf a-b}),  i.e. the  mean firing  rate  and the
variance  of  the activity  (computed  on  instantaneous firing  rates
evaluated with a $50$ ms Gaussian filter).

For low  couplings strengths, the  mean firing-rate in the  network is
close   to   the   value    predicted   for   the   fixed   point   of
Eq.~\eqref{eq:rate-lif1}, i.e.  the  equilibrium asynchronous state, and
essentially  independent  of the  refractory  period. Similarly,  the
variance  of the  activity remains  at low  values independent  of the
refractory period.   As the synaptic  strength is increased,  the mean
firing   rate   deviates  positively   from   the  equilibrium   value
(Fig.~\ref{fig:spk}  {\bf  a}),  and  the  variance  of  the  activity
increases (Fig.~\ref{fig:spk} {\bf b}). For intermediate and strong synaptic coupling, the values of first- and second-order activity statistics become dependent on the values of the refractory period.

Specifically,  for  intermediate  values of  the  coupling,  the
mean-firing  rate  increases with  decreasing  refractory period,  but
saturates with  decreasing refractory period  (Fig.~\ref{fig:spk} {\bf
	e}).   This  is similar to the  behavior of the rate  networks in the
inhibition-stabilized  fluctuating regime.  For large  values  of the
coupling,  the mean-firing  rate  instead diverges  linearly with  the
inverse  of  the refractory  period  (Fig.~\ref{fig:spk}  {\bf f}),  a
behavior analogous  to rate networks in the  second fluctuating regime
in which  the dynamics are only  stabilized by the upper  bound on the
activity.  The strength of the sensitivity to the refractory period depends on the inhibitory coupling: the stronger the relative inhibitory coupling, the weaker the sensitivity to the refractory period (Fig.~\ref{fig:spk}  {\bf d}).

The main  qualitative signatures of the two  fluctuating regimes found
in networks of  rate units are therefore also  observed in networks of
spiking LIF  neurons. It should be  however noted that  the details of
the dynamics  are different in  rate and LIF networks.  In particular,
the shape  of auto-correlation functions is different,  as LIF neurons
display  a richer  temporal structure   at low and intermediate coupling strengths. At strong coupling, the auto-correlation function resembles those of rate networks with spiking interactions (see Fig.~\ref{fig:noise} c), in particular it displays a characteristic cusp at zero time-lag. The simulated LIF networks show no sign of critical slowing down, as expected from the analysis of the effects of spiking noise on the activity.

Moreover, strong finite-size effects are  present in the simulations.  To
quantify correlations among units  and synchrony effects deriving from
finite-size  effects,  we  measure   the  standard  deviation  of  the
amplitude  of   fluctuations  in  the   population-averaged  activity,
normalized   by   the   square   root   of  the   mean   firing   rate
(Fig.~\ref{fig:spk} {\bf g}). Correlations  and synchrony appear to be
stronger  for small  values of  the refractory  period. The  effect of
correlations  is  furthermore weaker  in  the  low  and high  coupling
regimes, and  it has a  maximum for intermediate  couplings.  However,
whatever the value of $J$, they  decay as the system size is increased
(for a more detailed characterization, see \emph{Methods}).

In  summary,  for  the  range  of  values  of  the  refractory  period
considered here,  the activity in a  network of spiking  neurons is in
qualitative  agreement  with predictions  of  the  simple rate  models
analyzed  in  the previous  sections.  The  rate  model introduced  in
Eq.~\eqref{eq:rate-lif1}  however does  not  provide exact  quantitative
predictions for  the firing rate statistics above  the instability. In
particular, due to the  numerical limitations in considering the limit
$\tau_{rp}  \rightarrow 0$,  it is  not possible  to  evaluate exactly
through  simulations  the position  of  an  equivalent critical  value
$J_D$.

\newpage

\section*{Discussion}

We investigated  the fluctuating  dynamics of sparsely  connected rate
networks  with segregated  excitatory  and inhibitory  subpopulations.
We focused  on  a simplified  network architecture,  in
which  excitatory  and   inhibitory  neurons  receive  statistically
equivalent inputs,  but differ in  their output synaptic weights. In that  case, the
dynamical  mean field equations  that describe  the dynamics  can be
fully    analyzed.

Our central  result is that  in presence of excitation,  two different
regimes of  fluctuating activity appear as coupling  is increased. The
distinction between  these two regimes  rests on whether the  lower or
the  upper   bound  on   activity  stabilize  network   activity.   At
intermediate couplings, the fluctuating  activity is stabilized by the
lower bound  that enforces positive  firing rates, and  remains finite
even in absence  of upper bound.  For very  strong coupling, the upper
bound plays instead the dominant  role, as in its absence fluctuations
become unstable  and the network shows run-away  activity. This second
fluctuating  regimes is  absent in  purely inhibitory  networks  as it
requires excitatory feedback.

We also showed that in  presence of excitation, in  networks  with  fixed
in-degree,  self-generated fluctuations  strongly  affect first  order
statistics  such as  the  mean firing  rate,  which display  important
deviations  from values predicted  for the  fixed point  for identical
coupling strengths. Such deviations of mean firing rates are therefore
a  signature  of  underlying  fluctuations \cite{Ostojic}.  At  strong
coupling, in the second fluctuating  regime, both the first and second
order statistics monotonically increase with the upper bound. 

We solved  rigorously the DMF equations in  simplified networks, where
the  in-degree is  fixed  and excitatory  and  inhibitory neurons  are
statistically equivalent.   We showed however  that the classification
into three regimes extends to more general networks with statistically
distinguishable   populations   and   heterogeneous  in-degrees.    In
particular, signatures of the  two different fluctuating regimes are clearly apparent even
when  the  network  receives  strong  external  noise.  Finally  these
signatures  are also  seen in  networks of  integrate-and-fire neurons,
which display qualitatively similar dynamical features.

\subsection*{Relation to previous works}

The  transition from  fixed point  to fluctuating  activity  was first
studied by  Sompolinsky, Crisanti and  Sommers \cite{Sompolinsky}.  In
that classical work, the  connectivity was Gaussian and the activation
function  symmetric around  zero,  so that  the  dynamics exhibited  a
sign-reversal symmetry.  An important  consequence of this symmetry is
that  the  mean activity  was  always  zero,  and the  transition  was
characterized solely  in terms of second-order  statistics, which were
described through a dynamical mean field equation.

Recent studies  have examined more general  and biologically plausible
networks   \cite{Aljadeff1,Aljadeff2,Harish,Kadmon}.   Two   of  those
studies   \cite{Harish,Kadmon}  derived  dynamical   mean field  (DMF)
equations  for  networks  with  segregated  excitatory  and  inhibitory
populations,    and    asymmetric,    positively   defined    transfer
functions. The DMF  equations are however challenging to  solve in the
general  case of  two distinct  excitatory and  inhibitory populations
(see Methods). The two studies \cite{Harish,Kadmon} therefore analyzed
in detail  DMF solutions for purely inhibitory  networks, and explored
fluctuating activity in  excitatory-inhibitory networks mainly through
simulations.

In  contrast to  these recent  works, here  we exploited  a simplified
network architecture, in which DMF equations can be solved for
excitatory-inhibitory  networks. We found  the presence  of excitation
qualitatively  changes  the  nature   of  the  dynamics,  even  though
inhibition dominates. In  purely inhibitory networks, fluctuations are
weaker than in excitatory-inhibitory networks,  and as a result only weakly
affect   first-order  statistics.   

In \cite{Kadmon},  the authors  used transfer functions  without upper
bounds, and found that the chaotic state can undergo an instability in
which the  activity diverges. This instability is  directly related to
the transition between the  two fluctuating regimes which we studied in detail for
bounded  transfer functions.  Here we  showed that  these two
dynamical  regimes can in fact be distinguished  only if  the upper  bound is
varied: for  a fixed upper  bound, there is  no sign of  a transition.
Moreover, we showed that excitation  is  required for the appearance of
the second fluctuating  regime, as this regime relies on  positive feedback. For
purely  inhibitory networks,  in  which positive  feedback is  absent,
simulations show  that the second  fluctuating regime does  not occur,
although it  is predicted by  dynamical mean field theory:  indeed DMF
relies  on  a  Gaussian  approximation  which does  not  restrict  the
interactions  to  be strictly  negative,  and therefore  artifactually
introduces positive feedback at strong coupling.

The  previous studies  \cite{Harish,Kadmon} focused  on  networks with
random in-degree or Gaussian  coupling. In such networks, the quenched
component of  the coupling matrix  leads to quenched  heterogeneity in
the  stationary solution.   In  the present  work,  we instead  mostly
studied  networks with  fixed in-degree.   We  showed that  in such  a
setting a homogeneous distribution is the stable solution, so that the
quenched  variability is not required for the transition to fluctuating activity.

\subsection*{Synaptic timescales and rate fluctuations in networks of integrate-and-fire neurons}

Under which conditions a  regime analogous to rate  chaos develops in  networks of integrate-and-fire
neurons has  been a topic  of intense debate \cite{Ostojic,Engelken,Ostojic2015,Harish,Kadmon}. Two  different scenarios
have  been proposed:  (i) rate  chaos develops  in networks  of spiking
neurons  only  in  the  limit   of  very  slow  synaptic  or  membrane
time-constants  \cite{Harish,Kadmon};  (ii)   rate  chaos  can develop  in
generic  excitatory-inhibitory  networks,  i.e. for  arbitrarily  fast
synaptic time-constants  \cite{Ostojic}. The  heart of the  debate has
been  the nature of the signature of  rate chaos in spiking networks.

The  classical signature  of  the  transition to  rate  chaos is  {\em
	critical  slowing-down}, i.e.   the divergence  of the  timescale of
rate       fluctuations       close       to      the       transition
\cite{Sompolinsky,Engelken}.   Importantly,  this  signature   can  be
observed  only  if  the  coupling   is  very  close  to  the  critical
value. Moreover, as shown in \cite{Kadmon,Goedeke2016}, and reproduced
here (Fig.~\ref{fig:noise}), spiking  interactions induce noise in the
dynamics, and critical slowing down is very sensitive to the amplitude
of  such noise.   The  amplitude  of this  spiking  noise is  moreover
proportional  to  $1/\sqrt{\bar{\tau}}$,  where  $\bar{\tau}$  is  the
timescale  of  the rate  model,  usually  interpreted  as the  slowest
timescale    in   the    system   (either    membrane    or   synaptic
timescale). Critical-slowing down can  therefore be observed only when
the membrane or  synaptic timescales are very slow  and filter out the
spiking noise \cite{Harish,Kadmon}.

Here we have shown that for networks with EI connectivity and positive
firing rates, a novel signature of fluctuating activity appears simply
at the level of mean and variance of firing-rates, which become highly
sensitive  to the  upper bound  at  strong coupling.   In contrast  to
critical slowing-down, this signature of strongly fluctuating activity
manifests  itself in  a large  range of  couplings above  the critical
value.  A  second difference with  critical slowing down is  that this
signature  of  fluctuating  activity  is  very robust  to  noise,  and
therefore  independent of the  timescale of  the synapses  or membrane
time constant.  Simulations  of networks of integrate-and-fire neurons
reveal  such   signatures  of  underlying   fluctuating  activity  for
arbitrarily fast synaptic  time-constants, although there is no sharp transition
in  terms of  critical  slowing down.

The  results  presented  here  therefore reconcile  the  two  proposed
scenarios.    A  sharp   phase-transition   to  fluctuating   activity
characterized by  critical slowing down  appears only in the  limit of
very slow  synaptic or membrane time-constants.   For arbitrarily fast
synaptic  time-constants,  there is  no  sharp  phase transition,  but
instead  a smooth  cross-over  to strongly  fluctuating activity  that
manifests itself  at larger couplings through high  sensitivity to the
upper bound of the activity.

\subsection*{Mean-field theories and rate-based descriptions of integrate-and-fire networks}

The dynamical mean field theory used here to analyze rate networks should be contrasted with mean field theories developed for integrate-and-fire networks.
Classical  mean  field  theories  for networks  of  integrate-and-fire
neurons  lead to  a  self-consistent firing  rate  description of  the
equilibrium  asynchronous state \cite{AmitBrunel,Brunel99,Brunel2000},
but this effective description is  however not consistent at the level
of the second order statistics. Mean field theories for IF neurons assume indeed that
the input  to each  neuron consists of  white noise,  originating from
Poisson spiking; however the firing of an integrate-and-fire neuron in
response   to   white-noise  inputs   is   in   general  not   Poisson
\cite{Ostojic2011},   so   that   the   Poisson  assumption   is   not
self-consistent.  In  spite of this,  mean field theory  predicts well
the   first-order   statistics    over   a   large   parameter   range
\cite{Grabska2014}, but fails at  strong coupling when the activity is
strongly non-Poisson \cite{Ostojic}.

Extending mean field  theory to determine analytically self-consistent
second-order statistics is  challenging for spiking networks.  Several
numerical         approaches        have         been        developed
\cite{Lerchner2006,Dummer2014,   Wieland2015},  but  their   range  of
convergence appears to be limited.  A recent analysis of that type has
suggested  the  existence of  an  instability  driven by  second-order
statistics as the coupling is increased \cite{Wieland2015}.

A simpler  route to incorporate non-trivial second  order statistics in
the mean  field description  is to describe  the different  neurons as
Poisson processes with rates that vary  in time. One way to do this is
to  replace  every  neuron   by  a  linear-nonlinear  (LN)  unit  that
transforms its inputs  into an output firing rate,  and previous works
have shown that such an  approximation can lead to remarkably accurate
results \cite{OstojicBrunel2011,Tetzlaff2012,Pernice2012,Shaffer2013}.
If one  moreover approximates the linear  filter in the LN  unit by an
exponential,  this approach  results in  a mapping  from a  network of
integrate-and-fire neurons  to a network of rate  units with identical
connectivity \cite{Ostojic}.  Note that  such an approximation  is not
quantitatively  accurate for  the leaky  integrate-and-fire model with fast synaptic timescales  -  indeed the
linear  response  of  that   model  contains  a  very  fast  component
($1/\sqrt{t}$ divergence  in the impulse response at  short times, see
\cite{OstojicBrunel2011}).   A  single  timescale  exponential  however
describes  much   better  dynamics  of  other  models,   such  as  the
exponential integrate-and-fire \cite{OstojicBrunel2011}. The accuracy of the mapping from integrate-and-fire to rate networks also depends on synaptic timescales which influence both the amplitude of synaptic noise and the transfer function itself \cite{Brunel2001}. It has been argued that the mapping becomes exact in the limit of infinitely long timescales \cite{Shriki, Harish}.

In this  study, we  have analyzed rate  networks using  dynamical mean
field theory.  This version of mean field theory is different from the
one   used   for   integrate-and-fire   networks  as   it   determines
self-consistently   and   analytically   not  only   the   first-order
statistics,  but  also  the  second-order statistics,  i.e.  the  full
auto-correlation  function  of  neural  activity. Note  that  this  is
similar  in spirit  to the  approach developed  for integrate-and-fire
networks  \cite{Lerchner2006,Dummer2014,   Wieland2015},  except  that
integrate-and-fire  neurons  are  replaced  by  simpler,  analytically
tractable  rate units.  Dynamical  mean field  theory reveals  that at
large coupling,  network feedback strongly  amplifies the fluctuations
in the  activity, which  in turn  lead to an  increase in  mean firing
rates,  as seen in  networks of  spiking neurons  \cite{Ostojic}.  The
rate-model moreover  correctly predicts that for  strong coupling, the
activity is highly sensitive to  the upper bound set by the refractory
period, although the mean activity is well below saturation.

As pointed out above, the mapping from an integrate-and-fire to a rate network is based on a number of approximations and simplifications. The fluctuating state in
the rate network  therefore does not in general lead to  a quantitatively correct
description   of   the activity   in    a   network   of
integrate-and-fire neurons.  However, the  rate model does capture the
existence of  a fundamental instability,  which amplifies fluctuations
through network feedback.

\section*{Methods}

\subsection*{Rate network model}

We investigate the dynamics of a rate network given by:
\begin{equation}
	\dot{x_i}(t)=-x_i(t)+\sum_{j=1}^{N} J_{ij} \phi (x_j(t)) + I
	\label{eq:dyn_mem}
\end{equation}
where the index $i$ runs over  the units of the network. Each variable
$x_i$ is  interpreted as  the total input  current to neuron  $i$, and
$\phi(x)$  is  a monotonic,  positively  defined activation  function,
transforming  currents into  output  firing rates.   $I$  is a  common
external  input  and $J_{ij}$  a  random  synaptic  matrix. We  have
rescaled time to set the  time constant  to unity. All quantities are therefore taken to be unitless.

We  consider a  two-population (excitatory  and  inhibitory), sparsely
connected  network. All  the  excitatory synapses  have strength  $J$,
while all  inhibitory synapses have strength $-gJ$,  the parameter $g$
playing  the   role  of  the   relative  amount  of   inhibition  over
excitation. In  the simplest model  we consider, each  neuron receives
exactly $C$ incoming connections, with $1 \ll C\ll N$ \cite{Brunel2000}. A fraction $f$
of  inputs  are  excitatory  ($C_E=fC$), the  remaining  are  inhibitory
($C_I=(1-f)C$). We set $f=0.8$.

For the sake of simplicity, in most of the applications we restrict ourself to the case of a threshold-linear activation function with an offset $\gamma$. For practical purposes, we take:

\begin{equation}
	\phi(x) = \begin{cases}
		0 &  x<-\gamma \\
		\gamma+ x  &  -\gamma \leq x \leq \phi_{max}-\gamma\\
		\phi_{max} &  x>\phi_{max}-\gamma  
	\end{cases} \label{eq:phi_mem}
\end{equation}
where $\phi_{max}$ plays the role of the saturation value. In the following, we set $\gamma=0.5$. In most of the applications, if not explicitly indicated, we consider networks with no external input, and set $I=0$.

\subsection*{Mean field theory derivation} 
Here we derive in detail the Dynamical Mean Field (DMF) equations for the simplest excitatory-inhibitory network where the number of incoming connections $C$ is identical for all units.
For networks of large size, mean field theory provides a simple effective description of the network activity. More specifically, here we consider the limit of large $N$ while $C$ (and synaptic strengths) are held fixed \cite{AmitBrunel,Brunel2000}. The derivation provided here follows the same steps as in \cite{Sompolinsky,Rajan}, but takes into account non-vanishing first moments.

The dynamics of the network  depend on the specific realization of the
random connectivity matrix. The evolution of the network can therefore
be  seen  as a  random  process, which  can  be  characterized by  its
moments, obtained  by averaging over realizations  of the connectivity
matrix. The  dynamics can  be described either  by the moments  of the
synaptic  currents   $x_i$,  or  by   moments  of  the   firing  rates
$\phi(x_i)$.  The two  sets of  moments  are coupled,  and DMF  theory
exploits a Gaussian approximation to  derive a closed set of equations
for the first- and second-order  moments. This closed set of equations
can then be solved self-consistently.

More specifically, DMF theory acts by replacing the fully deterministic coupling term $\sum_{j} J_{ij} \phi (x_j)+I$ in Eq.~\eqref{eq:dyn_mem} by an equivalent Gaussian stochastic process $\eta_i$.
The effective mean field dynamics are therefore given by:
\begin{equation}
	\dot{x_i}(t)=-x_i(t)+\eta_i(t)
	\label{eq:langevin_mem}
\end{equation}
where the distribution of $\eta_i$ should effectively mimic the statistics of the original system in Eq. \eqref{eq:dyn_mem}. 

To be able to compute the moments of the synaptic currents $x_i$ and firing rates $\phi(x_i)$, the first step is to  compute self-consistently the first and second order moments of the effective noise $\eta_i$. 
For this,  averages over units, initial conditions, time and synaptic matrix instances (that we will indicate with $\langle\rangle$) are substituted with averages over the distribution of the stochastic process (that we will indicate with $[]$). For the mean, we get:
\begin{equation}
	\begin{split}
		&[\eta_i(t)]=\langle\sum_{j=1}^{N} J_{ij} \phi (x_j) + I\rangle=\sum_{j_E=1}^{C_E}J\langle\phi(x_{j_E})\rangle - g \sum_{j_I=1}^{C_I}J\langle\phi(x_{j_I})\rangle + I\\
		&=J(C_E-gC_I)\langle\phi\rangle + I
	\end{split}
\end{equation}
where the indices $j_E$ and $j_I$ run over the excitatory and the inhibitory units pre-synaptic to unit $i$.

Following previous works \cite{Sompolinsky, Cessac}, here we assume that, for large $N$, $J_{ij}$ and $\phi (x_j)$ behave independently. Moreover, we assume that the mean values of $x$ and $\phi$  reach stationary values for $t\rightarrow \infty$, such that $[\eta_i(t)]=[\eta_i]$. 

Under the same hypothesis, the second moment $[\eta_i(t)\eta_j(t+\tau)]$ is given by:
\begin{equation}
	[\eta_i(t)\eta_j(t+\tau)]=\langle\sum_{k=1}^{N} J_{ik} \phi (x_k(t))\sum_{l=1}^{N} J_{jl} \phi (x_l(t+\tau))\rangle+2I J(C_E-gC_I)\langle\phi\rangle+I^2.
\end{equation}

In order to evaluate the first term in the r.h.s., we differentiate two cases: first, we take $i=j$, yielding the noise auto-correlation.
We assume that in the thermodynamic limit, where the Langevin equations in Eq.~\eqref{eq:langevin_mem} decouple, different units behave independently: $\langle\phi(x_k) \phi (x_l)\rangle=\langle\phi(x_k)\rangle \langle\phi (x_l)\rangle$ if $k \neq l$. We will verify this assumption self-consistently by showing that, in the same limit, $[\eta_i(t)\eta_j(t+\tau)] =0 $ when $i \neq j$.

The sum over $k$ ($l$) can be split into a sum over $k_E$ and $k_I$ ($l_E$ and $l_I$) by segregating the contributions from the two populations. We thus get:
\begin{equation}
	\begin{split}
		&\langle\sum_{k=1}^{N} J_{ik} \phi (x_k(t))\sum_{l=1}^{N} J_{il} \phi (x_l(t+\tau))\rangle=\langle\sum_{k_E=1}^{N_E} J_{ik_E} \phi (x_{k_E}(t))\sum_{l_E=1}^{N_E} J_{il_E} \phi (x_{l_E}(t+\tau))\rangle\\
		&+ \langle\sum_{k_I=1}^{N_I} J_{ik_I} \phi (x_{k_I}(t))\sum_{l_I=1}^{N_I} J_{il_I} \phi (x_{l_I}(t+\tau))\rangle+ 2 \langle\sum_{k_E=1}^{N_E} J_{ik_E} \phi (x_{k_E}(t))\sum_{l_I=1}^{N_I} J_{il_I} \phi (x_{l_I}(t+\tau))\rangle.
	\end{split}
\end{equation}
We focus on the first term of the sum (same considerations hold for the second two), and we differentiate contributions from $k_E=l_E$ and $k_E\neq l_E$. Setting $k_E=l_E$ returns a contribution equal to $C_EJ^2\langle\phi^2\rangle$. In the sum with $k_E \neq l_E$, as $C$ is fixed, we obtain exactly $C_E(C_E-1)$ contributions of value $J^2\langle \phi \rangle^2$. This gives, for the two populations:
\begin{equation}
	\begin{split}
		&\langle\sum_{k=1}^{N} J_{ik} \phi (x_k(t))\sum_{l=1}^{N} J_{il} \phi (x_l(t+\tau))\rangle=C_E J^2\langle\phi(x_i(t))\phi(x_i(t+\tau))\rangle+C_E(C_E-1)J^2\langle\phi\rangle^2\\
		&-2C_E C_I gJ^2\langle\phi\rangle^2+C_Ig^2 J^2 \langle\phi(x_i(t))\phi(x_i(t+\tau))\rangle+ C_I(C_I-1)g^2 J^2 \langle \phi\rangle^2\\
		& = J^2(C_E+g^2C_I)\langle\phi(x_i(t))\phi(x_i(t+\tau))\rangle + J^2(C_E-gC_I)^2\langle\phi\rangle^2 -  J^2(C_E+g^2C_I) \langle\phi\rangle^2. 
	\end{split}
\end{equation}

By defining the rate auto-correlation function $C(\tau)=\langle\phi(x_i(t))\phi(x_i(t+\tau))\rangle$, we finally get:
\begin{equation}
	[\eta_i(t)\eta_i(t+\tau)]-[\eta_i]^2=J^2(C_E+g^2C_I)\{ C(\tau) - \langle \phi\rangle^2\}.
	\label{eq:etasq}
\end{equation}

When $i \neq j$,  we instead obtain:
\begin{equation}
	\begin{split}
		&\langle\sum_{k=1}^{N} J_{ik} \phi (x_k(t))\sum_{l=1}^{N} J_{jl} \phi (x_l(t+\tau))\rangle= C_E^2 J^2 \langle\phi\rangle^2 + p C_E J^2 \{ C(\tau)- \langle\phi\rangle^2\}+C_I^2 g^2 J^2 \langle\phi\rangle^2 \\
		& +p C_I g^2 J^2 \{ C(\tau)-\langle\phi\rangle^2\} - 2 C_E C_I g J^2 \langle\phi\rangle^2.
	\end{split}
\end{equation}
The constant $p$ corresponds to the probability that, given that $k$ is a pre-synaptic afferent of neuron $i$, the same neuron is connected also to neuron $j$. Because of sparsity, we expect this value to be small. More precisely, since $N$ is assumed to be large, we can approximate the probability $p$ with $C/N$.
We eventually find:
\begin{equation}
	[\eta_i(t)\eta_j(t+\tau)]-[\eta_i][\eta_j]=pJ^2(C_E+g^2C_I)\{ C(\tau) - \langle\phi\rangle^2\} \to 0
\end{equation}
because $p \rightarrow 0$ when $N \rightarrow \infty$.

Once the statistics of the effective stochastic term $\eta_i$ are known, we can  describe  the input current $x$ in terms of its mean $\mu=[x_i]$ and its mean-subtracted correlation function $\Delta(\tau)= [x_i(t) x_i(t+\tau)] - [x_i]^2$. 
The mean field current $x_i(t)$ emerging from the stochastic process in Eq.~\eqref{eq:langevin_mem} behaves as a time-correlated Gaussian variable. First we observe that, asymptotically, its mean value $\mu$  coincides with the mean of the noise term $\eta_i$:
\begin{equation}
	\mu= J(C_E-gC_I)[\phi]+I
\end{equation}

By differentiating twice $\Delta(\tau)$ with respect to $\tau$ and using equations (\ref{eq:langevin_mem}) and (\ref{eq:etasq}), we moreover get a second-order differential equation for the auto-correlation evolution:
\begin{equation}
	\ddot{\Delta}(\tau)=\Delta(\tau)-J^2(C_E+g^2C_I)\{C(\tau)-\langle\phi\rangle^2\}.
	\label{eq:deltaev_mem}
\end{equation} 

By explicitly constructing $x(t)$ and $x(t+\tau)$ in terms of unit Gaussian variables, we self-consistently rewrite the firing rate statistics $[\phi]$ and $C(\tau)$, as integrals over the Gaussian distributions:
\begin{equation}
	\begin{split}
		&[\phi]=\int \mathcal{D}z \phi(\mu + \sqrt{\Delta_0}z)\\
		& C(\tau)=\int \mathcal{D}z \left[ \int \mathcal{D}y \phi(\mu +\sqrt{\Delta_0-\Delta(\tau)} y + \sqrt{\Delta(\tau)}z) \right]^2  \label{eq:c_explicit}
	\end{split}
\end{equation}
where we used the short-hand notation: $\int \mathcal{D}z = \int_{-\infty}^{+\infty} \frac{e^{-\frac{z^2}{2}}}{\sqrt{2 \pi}} dz$. For reasons which will become clearly soon, we can focus on positive values of the auto-correlation $\Delta$. We moreover defined  $\Delta_0=\Delta(\tau=0)$.

Following \cite{Sompolinsky}, Eq.~\eqref{eq:deltaev_mem} can be seen as  analogous to the equation of motion of a classical particle in a one-dimensional potential:
\begin{equation}
	\ddot{\Delta}=-\frac{\partial V(\Delta, \Delta_0)}{\partial \Delta}.
\end{equation}
The potential $V(\Delta, \Delta_0)$  can be derived by integrating the
right-hand side of  Eq.~\eqref{eq:deltaev_mem} over $\Delta$ and using
Eq.~\eqref{eq:c_explicit}. This yields
\begin{equation}
	V(\Delta, \Delta_0)=-\frac{\Delta^2}{2}+J^2(C_E+g^2 C_I) \left\{ \int \mathcal{D}z  \left[ \int \mathcal{D}y \Phi (\mu+\sqrt{\Delta_0-\Delta}y+\sqrt{\Delta}z) \right]^2 - \Delta[\phi]^2\right\}
	\label{eq:V_mem}
\end{equation}
where $\Phi(x)=\int_{-\infty}^x dz \phi(z)$. 

In absence of external noise, the initial condition to be satisfied is $\dot{\Delta}(\tau=0)=0$, which implies null kinetic energy for $\tau=0$. A second condition is given by: $\Delta_0>|\Delta(\tau)|$ $\forall \tau$.
The solution $\Delta(\tau)$ depends on the initial value $\Delta_0$, and it is governed by the energy conservation law:
\begin{equation}
	V(\Delta(\tau=0), \Delta_0)=V(\Delta(\tau=\infty), \Delta_0)+\frac{1}{2}\dot{\Delta}(\tau=\infty)^2.
	\label{eq:conservation_mem}
\end{equation}

The stationary points and the qualitative features of the $\Delta(\tau)$ trajectory depend then on the shape of the potential $V$.
We notice that the derivative of the potential in $\Delta=0$ is always 0, suggesting a possible equilibrium point where the current  distribution is concentrated in its mean value $\mu$. Note that the existence of the stationary point in 0  stems from the $-\Delta [\phi]^2$ term in the potential, which comes from taking the connectivity degree $C$ fixed for each unit in the network (for a comparison with the equations obtained for random in-degree networks, see below).

\begin{figure}[h!]
	\begin{adjustwidth}{-0.3in}{-0.3in} 
		\centering
		\includegraphics{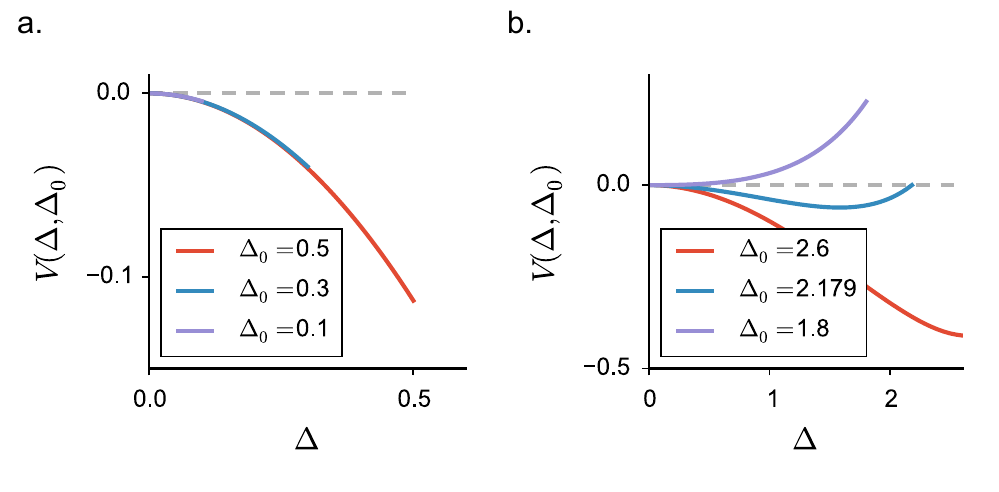}
		\caption{{\bf Dynamical mean field potential $V(\Delta, \Delta_0)$ for different values of the parameter $\Delta_0$}; fixed $\mu$. The activation function is chosen to be threshold-linear. \textbf{a.} $J<J_C$: the potential is always concave. \textbf{b.} $J>J_C$: the shape of the potential strongly depends on the value of $\Delta_0$.  }
		\label{fig:potential}
	\end{adjustwidth}
\end{figure}

When the first moment
$\mu$ is determined self-consistently, the shape of $V$ depends on the values of $J$ and $\Delta_0$ (Figure~\ref{fig:potential} {\bf a-b}). In particular, a critical value $J_C$ exists such that:
\begin{itemize}
	\item when $J<J_C$, the potential has the shape of a concave parabola centered in $\Delta=0$ (Figure~\ref{fig:potential} {\bf a}). The only physical bounded solution is then $\Delta=\Delta_0=0$;
	\item when $J>J_C$, the potential admits different qualitative configurations and an infinite number of different $\Delta(\tau)$ trajectories. In general, the motion in the potential will be oscillatory (Figure~\ref{fig:potential} {\bf b}).
\end{itemize}

However, in the strong coupling regime, a particular solution exists, for which $\Delta(\tau)$ decays to 0 as $\tau \rightarrow \infty$. In this final state, there is no kinetic energy left. A monotonically decaying auto-correlation function is the only stable solution emerging from numerical simulations.

For this particular class of solutions, eq. (\eqref{eq:conservation_mem}) reads:
\begin{equation}
	V(\Delta_0, \Delta_0)=V(0, \Delta_0).
\end{equation}
More explicitly:
\begin{equation}
	\begin{split}
		\frac{\Delta_0^2}{2} =  J^2(C_E+g^2C_I)  & \left\{  \int \mathcal{D}z \Phi^2 (\mu+\sqrt{\Delta_0}z) - \left( \int \mathcal{D}z \Phi (\mu+\sqrt{\Delta_0}z) \right)^2 \right.\\
		& \left. -\Delta_0 \left( \int \mathcal{D}z \phi (\mu+\sqrt{\Delta_0}z) \right)^2  \right\}
	\end{split}
\end{equation}
which gives an equation for $\mu$ and $\Delta_0$ to be solved together with the equation for the mean:
\begin{equation}
	\mu=J(C_E-gC_I)\int \mathcal{D}z \phi(\mu + \sqrt{\Delta_0}z)+I
\end{equation}

In a more compact form, we can reduce the system of equations to:
\begin{equation}
	\begin{split}
		\mu=& J(C_E-gC_I)[\phi]+I\\
		\frac{\Delta_0^2}{2}=& J^2(C_E+g^2C_I)\left\{[\Phi^2]-[\Phi]^2 -\Delta_0[\phi]^2\right\}.
	\end{split}
	\label{eq:DMF}
\end{equation}
Once $\mu$ and $\Delta_0$ are computed, their value can be injected into equation (\ref{eq:deltaev_mem}) to get the time course of the auto-correlation function.

Not surprisingly, the results above rely on the assumption of sparsity in the connectivity: $C \ll N$. Classic DMF theory, indeed, requires synaptic entry $J_{ij}$ to be independent one from each other. Fixing the number of non-zero connections for each unit is imposing a strong dependence among the entries in each row of the synaptic matrix.
Nevertheless, we expect this dependence to become very weak when $N\rightarrow \infty$, and we find that DMF can still predict correctly the system behavior, keeping however a trace of the network homogeneity through the term $-[\phi]^2$ in Eq.~\eqref{eq:deltaev_mem}.
Fixing the degree $C$ sets to zero the asymptotic value of the auto-correlation function, and results in a perfect self-averaging and homogeneity of activity statistics in the population.

To conclude, we note that finding the DMF solution for an excitatory-inhibitory network reduces here to solving a  system of two-equations. A large simplification in the problem comes here from considering networks where excitatory and inhibitory units receive statistically equivalent inputs.
DMF theory models indeed the statistical distribution of the input currents inside each network unit. For this reason, it does not include any element deriving from the segregation of the excitatory and the inhibitory populations in a two-columns connectivity structure. In consequence, for identical sets of parameters, we expect the same DMF equations to hold in more generic networks, where each neuron receive $C_E$ excitatory and $C_I$ inhibitory inputs, but can make excitatory or inhibitory output connections. We checked the validity of this observation (see later in \emph{Methods}).

In a more general case, where excitation and inhibition are characterized as distinguishable populations with their own statistics, solving the DMF equations becomes  computationally costly. The main complication comes from the absence of any equivalent classical motion in a potential. For that reason, previous studies have focused mostly on the case of purely inhibitory populations \cite{Kadmon, Harish}.

\paragraph*{Second critical coupling $J_D$}
While  the DMF  equations  can  be derived  for  a generic  activation
function $\phi(x)$,  here we  focus, for mathematical  convenience, on
the  simple  case  of  the  threshold-linear  activation  function  in
Eq.~\eqref{eq:phi_mem}.   From  now  on,  for simplicity,  we  will  set
$I=0$.  For  each value of the synaptic  strength $J$, the
system  in   Eq.~\eqref{eq:DMF}  allows   to  compute  the   first-  and
second-order  statistics   of  the  network  activity.   As  shown  in
\emph{Results}, DMF revels the  existence of two different fluctuating
regimes  above  the  critical  coupling  $J_C$, governed  by  the  two
different non-linear  constraint in  the dynamics: the  positivity and
the saturation of firing rates.

Here we  study the behavior of  the DMF solution close  to the second
critical coupling  $J_D$, in the  case of a  non-saturating activation
function  where $\phi_{max}\rightarrow  \infty$.  When $J$  approaches
$J_D$, $\Delta_0 \rightarrow  \infty$, while $\mu \rightarrow -\infty$
(Fig.~\ref{fig:tworeg}).

Led by dimensionality arguments, we assume that, close to the divergence point, the ratio $k=\mu/\sqrt{\Delta_0}$ is  constant. With a threshold-linear transfer function, it is possible to compute analytically the three Gaussian integrals implicit in (\ref{eq:DMF}) and to provide an explicit analytic form of the DMF equations. The equation for the mean translates into:
\begin{equation}
	\mu=J(C_E-gC_I) [\phi]=J(C_E-gC_I)\left\{\left(\frac{1}{2}+\mu \right) \left( \frac{1}{2} -g (x_a) \right) + \sqrt{\frac{{\Delta_0}}{{2 \pi}}}e^{-\frac{1}{2}x_a^2}\right\}.
\end{equation}
where $x_a=\frac{1}{\sqrt{\Delta_0}}\left(-\frac{1}{2}-\mu \right)\sim -k$ and where we have defined: $g(x)=\frac{1}{2} \erf(x/\sqrt{2})$. When $J \rightarrow J_D$, by keeping only the leading order in $\sqrt{\Delta_0}$, we find $\mu=\hat{k}\sqrt{\Delta_0}$ with:
\begin{equation}
	\hat{k}=\frac{J(C_E-gC_I)\frac{e^{-\frac{k^2}{2}}} {\sqrt{2 \pi}} }{1-J(C_E-gC_I)(\frac{1}{2}+G(k))}.
\end{equation}

By imposing $k=\hat{k}$, one can determine self-consistently  the value of $k$ for each value of $J$. We introduce $\mu=k\sqrt{\Delta_0}$ into the second equation for $\Delta_0$. By keeping only the leading order in $\Delta_0$, we find:
\begin{equation}
	\begin{split}
		&\sqrt{\Delta_0}=f(k)\\
		f&(k)=\frac{J^2(C_E+g^2C_I) T(k)}{\frac{1}{2}-J^2(C_E+g^2C_I) S(k)}
	\end{split}
	\label{eq:jd_mem}
\end{equation}
with:
\begin{equation}
	\begin{split}
		S(k)=&\frac{1}{4}k^4\left[ \frac{1}{2} +g (k) \right]+\frac{1}{4}k^3\frac{e^{-\frac{k^2}{2}}} {\sqrt{2 \pi}}+k^2\left[\frac{3}{2}\left(\frac{1}{2}+ g(k)\right) -\left(\frac{1}{2}+ g(k)\right)^2\right]\\ &+k\left[\frac{5}{4}\frac{e^{-\frac{k^2}{2}}} {\sqrt{2 \pi}} -2\left( \frac{1}{2}+ g(k) \right) \frac{e^{-\frac{k^2}{2}}} {\sqrt{2 \pi}} \right]+ \frac{3}{4} \left(\frac{1}{2} +g (k) \right)- \left(\frac{e^{-\frac{k^2}{2}}} {\sqrt{2 \pi}} \right)^2\\
		& -\left\{\left(\frac{1}{2}k^2+\frac{1}{2}\right)\left[ \frac{1}{2} +g (k) \right]+\frac{1}{2}k \frac{e^{-\frac{k^2}{2}}} {\sqrt{2 \pi}}\right\}^2
	\end{split}
\end{equation}
In order to obtain a solution $\Delta_0$, from Eq.~\eqref{eq:jd_mem} we require the function $f(k)$ to be positive. We observe that $f$ diverges  when its denominator crosses zero.  Here $f(k)$ changes sign, becoming negative. We use this condition to determine $J_D$:
\begin{equation}
	J_D^2(C_E+g^2C_I)S(k(J_D))=\frac{1}{2}
\end{equation}

In presence of external noise of variance $2\Delta_{ext}$, the equation for $\Delta_0$ is perturbed by an additive term proportional to $\Delta_{ext}^2$ (see below in \emph{Methods}). Since we treat the noise variance as a constant, this additional term does not contribute to the divergence in the leading order in $\Delta_0$ (namely, $\Delta_0^{3/2}$, $\Delta_0^{2}$), and the presence of noise does not influence the value of $J_D$.

Similarly, when we add white noise to mimic the noise introduced by Poisson spikes, we find the extra term to be proportional to $[\phi]^2$, which is of the same order of $\Delta_0$. As a consequence, again, it does not perturb the equation for $\Delta_0$ to the leading orders.

\paragraph*{Mean field theory in presence of  noise}
In order to investigate the effect of an external noisy input on the dynamical regimes, we introduced an additive, white noise term in Eq.~\eqref{eq:dyn_mem}. The network dynamics in this case read:
\begin{equation}
	\dot{x_i}(t)=-x_i(t)+\sum_{j=1}^{N} J_{ij} \phi (x_j(t)) + \xi_i(t)
\end{equation}
with $\langle \xi_i(t) \rangle =0$ and $\langle \xi_i(t)\xi_j(t+\tau) \rangle =2 \Delta_{ext}\delta_{ij} \delta(\tau)$.

As above, we replace the forcing term $\sum_j J_{ij}\phi(x_j)+\xi_i$ by an effective noise $\eta_i$. By following the same steps as before we find:
\begin{equation}
	\begin{split}
		& [\eta_i(t)]=J(C_E-gC_I)\langle \phi \rangle\\
		&[\eta_i(t)\eta_i(t+\tau)]-[\eta_i]^2=\delta_{ij} \left[J^2 (C_E+g^2C_I) \left\{ C(\tau) -\langle \phi \rangle^2 \right\}+2 \Delta_{ext} \delta(\tau)\right]
	\end{split}
\end{equation}
which translates into:
\begin{equation}\label{eq:mfeq_noise}
	\ddot{\Delta}(\tau)=\Delta(\tau)-J^2(C_E+g^2C_I)\{C(\tau)-[\phi]^2\}+2 \Delta_{ext} \delta(\tau)
\end{equation} 

We conclude that the external noise acts on the auto-correlation function by modifying its initial condition into: $\dot{\Delta}(0^+)=-\dot{\Delta}(0^-)=-\Delta_{ext}$. In terms of the analogy with the 1D motion, the presence of noise translates into an additive kinetic term in $\tau=0$, which one has to take into account while writing down the energy balance:
\begin{equation}
	V(\Delta_0, \Delta_0)+\frac{1}{2}\dot{\Delta}(0)^2=V(0, \Delta_0)
\end{equation}
to be solved again together with the equation for the mean $\mu$. The potential $V(\Delta, \Delta_0)$, in contrast, remains unperturbed. 
The main effect of including a kinetic term at $\tau=0$ consists in allowing a variance $\Delta_0 \neq 0$ also in the low coupling regime, where the potential has the usual shape as in Fig.~\ref{fig:potential}(a).

From a mean field perspective, white noise can be studied as a proxy for the effect induced by spikes on the rate dynamics.
In order to better quantify this effect, following \cite{Kadmon}, we add a spiking mechanism on the rate dynamics in Eq.~\eqref{eq:dyn}. Spikes are emitted according to independent inhomogeneous Poisson processes of rate $\phi(x_j(t))$, which obeys:
\begin{equation}
	\bar{\tau}\dot{x}(t)=-x(t)+\sum_{j=1}^{N} J_{ij} \chi_j(t)  \label{eq:rate-spk}
\end{equation}
and $\chi_j(t)$ is the spike train emitted by neuron $j$: $\chi_j(t)=\sum_k\delta(t-t_j^k)$.

This simple spiking mechanism can be again incorporated into a DMF description. Here, following \cite{Kadmon}, we show that the resulting equations correspond to an usual rate model with additive white noise, whose variance is given by $J^2(C_E+g^2C_I)[\phi]/\bar{\tau}$.
The mean field forcing noise is in this case $\eta_i(t)=\sum_j J_{ij} \chi_j(t)$. By separating $J_{ij}$ into the sum of its mean and a zero-mean term, we get that the usual equation for the first order statistics holds:
\begin{equation}
	[\eta_i]=J(C_E-gC_I)[\phi]
\end{equation}

In order to compute the noise auto-correlation, we separate $\eta_i$ into a rate and a zero-mean spikes contribution: $\eta_i=\eta_i^{r}+\eta_i^{sp}$, where $\eta_i^{r}=\sum_j J_{ij} \phi(x_j)$ and $\eta_i^{sp}=\sum_j J_{ij} \{\chi_j-\phi(x_j)\}$. The auto-correlation of the rate component returns the usual contribution:
\begin{equation}
	[(\eta_i^{r}(t)-[\eta_i^{r}])(\eta_j^{r}(t+\tau)-[\eta_j^{r}])]=\delta_{ij}J^2(C_E+g^2C_I)\{C(\tau)-[\phi]^2\}
\end{equation}

while the auto-correlation of the spikes term generates the instantaneous variability induced by the Poisson process: 
\begin{equation}
	[(\eta_i^{sp}(t)-[\eta_i^{sp}])(\eta_j^{sp}(t+\tau)-[\eta_j^{sp}])]=\delta_{ij}J^2(C_E+g^2C_I)[\phi]\delta(\tau)
\end{equation}
By summing the two contributions together, and rescaling time appropriately, we obtain the evolution equation for $\Delta(\tau)$ equivalent to Eq. \eqref{eq:mfeq_noise} with a self-consistent white noise term:
\begin{equation}
	\ddot{\Delta}(\tau)=\Delta(\tau)-J^2(C_E+g^2C_I)\{C(\tau)-[\phi]^2 + \frac{[\phi]}{\bar{\tau}}\delta(\tau)\}
\end{equation} 

\paragraph*{Mean field theory in general EI networks}
We discuss here the more general case of a block connectivity matrix, corresponding to one excitatory and one inhibitory population receiving statistically different inputs. The synaptic matrix is now given by:
\begin{equation}
	\mathrm{J}=J\left(
	\begin{array}{c|c}
		\mathrm{J_{EE}}  &  \mathrm{J_{EI}} \\ \hline
		\mathrm{J_{IE}} &  \mathrm{J_{II}}
	\end{array}\right)
\end{equation}
Each  row of  $\mathrm{J}$  contains exactly
$C_E$  non-zero excitatory  entries in  the blocks  of  the excitatory
column,  and  exactly  $C_I$  inhibitory  entries  in  the  inhibitory
blocks. Non-zero elements are  equal to $j_E$ in $\mathrm{J_{EE}}$, to
$-g_Ej_E$ in $\mathrm{J_{EI}}$, to  $j_I$ in $\mathrm{J_{IE}}$, and to
$-g_Ij_I$  in $\mathrm{J_{II}}$. 

The network admits a fixed point $(x_0^E, x_0^I)$ which is homogeneous within the two different populations:
\begin{equation}
	\left(
	\begin{array}{c}
		x_0^E  \\
		x_0^I 
	\end{array}\right)=J
	\left(
	\begin{array}{c}
		j_E(C_E \phi(x_0^E)-g_E C_I \phi(x_0^I))  \\
		j_I(C_E \phi(x_0^E)-g_I C_I \phi(x_0^I))
	\end{array}\right)	
\end{equation}
With linear stability analysis, we obtain that the fixed point stability is determined by the eigenvalues of matrix:
\begin{equation}
	\mathrm{S}=J\left(
	\begin{array}{c|c}
		\phi'_E\mathrm{J_{EE}}  &  \phi'_I\mathrm{J_{EI}} \\ \hline
		\phi'_E\mathrm{J_{IE}} &  \phi'_I\mathrm{J_{II}}
	\end{array}\right)
\end{equation}
where we used the short-handed notation $\phi'_k=\phi'(x_0^k)$.

The eigenspectrum of $\mathrm{S}$ consists of a densely distributed component, represented by a circle in the complex plane, and a discrete component, consisting of two outlier eigenvalues.
The radius of the complex circle is determined by the $2\mathrm{x}2$ matrix containing the variance of the entries distributions in the four blocks, multiplied by $N$ \cite{Aljadeff1, Aljadeff2, AljadeffEig}:
\begin{equation}
	\mathrm{\Sigma}=J^2\left(
	\begin{array}{cc}
		{\phi' _E}^2 C_Ej_E^2  &  {\phi' _I} ^2 C_Ig_E^2j_E^2\\
		{\phi' _E} ^2 C_Ej_I^2 &  {\phi' _I} ^2 C_Ig_I^2j_I^2
	\end{array}\right)
\end{equation}
More precisely, the radius of the circle is given by the square root of its larger eigenvalues:
\begin{equation}
	\begin{split}
		r=& \left[\frac{1}{2}J^2 \right. \left\{ C_E{\phi'_E}^2j_E^2+C_I{\phi'_I}^2g_I^2j_I^2  \right.\\
		&\left. \left. + \sqrt{(C_E{ \phi'_E}^2j_E^2+C_I{\phi'_I}^2g_I^2j_I^2)^2-4C_EC_I{\phi'_E}^2 {\phi'_I}^2j_E^2j_I^2(-g_E^2+g_I^2)} \right\}\right]^{\frac{1}{2}}
	\end{split}
\end{equation}
where the derivative terms $\phi'^k$ contain an additional dependency on $J$.

In order to determine the two outlier eigenvalues,  we construct the  $2\mathrm{x}2$ matrix containing the mean of $\mathrm{S}$ in each of the four blocks, multiplied by $N$:
\begin{equation}
	\mathrm{M}=J\left(
	\begin{array}{cc}
		\phi'_E C_Ej_E  &  -\phi'_I C_Ig_Ej_E\\
		\phi'_E C_Ej_I &  -\phi'_I C_Ig_Ij_I
	\end{array}\right)
\end{equation}
The outliers correspond to the two eigenvalues of $\mathrm{M}$, and are given by:
\begin{equation}
	\xi_{\pm}=\frac{1}{2}J\left\{\phi'_E C_Ej_E-\phi'_IC_Ig_Ij_I \pm \sqrt{(\phi'_EC_Ej_E-\phi'_IC_Ig_Ij_I)^2+4\phi'_E \phi'_IC_EC_Ij_Ej_I(-g_E+g_I)} \right\}
\end{equation}

Notice that, if $g_E$ is sufficiently larger than $g_I$, the outlier eigenvalues can be complex conjugates.

We focus on the case where, by increasing the global coupling $J$, the instability to chaos is the first bifurcation to take place. As in the simpler case when excitatory and inhibitory populations are identical, we need the real part of the outliers to be negative or positive but smaller than the radius $r$ of the densely distributed component of the eigenspectrum. This requirement can be accomplished by imposing relative inhibitory strengths $g_E$ and $g_I$ strong enough to overcome excitation in the network.
For a connectivity matrix which satisfies the conditions above, an instability to a fluctuating regime occurs when the radius $r$ crosses unity.

We can use again DMF to analyze the network activity below the instability. To start with, dealing with continuous-time dynamics, one can easily generalize the mean field equations we recovered for the simpler two-column connectivity. In the new configuration, the aim of mean field theory is to determine two values of the mean activity and two values for the variance, one for each population. 

By following the same steps as before, we define $\eta_i^E=\sum_{j=1}^N J_{ij}\phi(x_j(t))$ for each $i$ belonging to the $E$ population, and $\eta_i^E=\sum_{j=1}^N J_{ij}\phi(x_j(t))$ for each $i$ belonging to $I$. Those two variables represent the effective stochastic inputs to excitatory or inhibitory units which replace the deterministic network interactions. Under the same hypothesis as before, we compute the statistics of the $\eta_i^E$ and $\eta_i^I$ distributions. For the mean, we find:
\begin{equation}
	\left(
	\begin{array}{c}
		\left[\eta_i^E\right]  \\
		\left[\eta_i^I\right]
	\end{array}\right)=J
	\left(
	\begin{array}{cc}
		C_Ej_E  &  -C_Ig_Ej_E\\
		C_Ej_I &  -C_Ig_Ij_I
	\end{array}\right)
	\left(
	\begin{array}{c}
		\left[\phi^E\right]  \\
		\left[\phi^I\right]
	\end{array}\right)
\end{equation}

For the second order statistics, we have: 
\begin{equation}
	\left(
	\begin{array}{c}
		\left[(\eta_i^E(t)-[\eta_i^E])(\eta_j^E(t+\tau)-[\eta_j^E])\right] \\
		\left[(\eta_i^I(t)-[\eta_i^I])(\eta_j^I(t+\tau)-[\eta_j^I])\right]
	\end{array}\right)=J^2
	\left(
	\begin{array}{cc}
		C_Ej_E^2  &  C_Ig_E^2j_E^2\\
		C_Ej_I^2 &  C_Ig_I^2j_I^2
	\end{array}\right)
	\left(
	\begin{array}{c}
		C^E(\tau)-\left[\phi^E\right]^2  \\
		C^I(\tau)-\left[\phi^I\right]^2
	\end{array}\right).
\end{equation}

By using those results, we obtain two equations for the mean values of the input currents:
\begin{equation}
	\left(
	\begin{array}{c}
		\mu^E  \\
		\mu^I
	\end{array}\right)=J
	\left(
	\begin{array}{cc}
		C_Ej_E  &  -C_Ig_Ej_E\\
		C_Ej_I &  -C_Ig_Ij_I
	\end{array}\right)
	\left(
	\begin{array}{c}
		\left[\phi^E\right]  \\
		\left[\phi^I\right]
	\end{array}\right).
	\label{eq:ei_1_mem}
\end{equation}
and two differential equations for the auto-correlation functions, which can be summarized as:
\begin{equation}
	\left(
	\begin{array}{c}
		\ddot{\Delta}^E(\tau) \\
		\ddot{\Delta}^I(\tau)
	\end{array}\right)=
	\left(
	\begin{array}{c}
		{\Delta}^E(\tau) \\
		{\Delta}^I(\tau)
	\end{array}\right)
	-J^2
	\left(
	\begin{array}{cc}
		C_Ej_E^2  &  C_Ig_E^2j_E^2\\
		C_Ej_I^2 &  C_Ig_I^2j_I^2
	\end{array}\right)
	\left(
	\begin{array}{c}
		C^E(\tau)-\left[\phi^E\right]^2  \\
		C^I(\tau)-\left[\phi^I\right]^2
	\end{array}\right).
	\label{eq:ei_2_mem}
\end{equation}
All the mean values are defined and computed as before, the population averages to be taken only over the $E$ or the $I$ population.

The main difficulty in solving Eqs.~\eqref{eq:ei_1_mem} and~\eqref{eq:ei_2_mem} comes from the absence of an analogy with an equation of motion for a classical particle in a potential. Unfortunately, indeed, isolating the self-consistent solution in absence of an analogous suitable potential $V(\Delta^E(\tau),\Delta^I(\tau))$ appears to be computationally very costly.

However, if we restrict ourselves to discrete-time rate dynamics:
\begin{equation}
	x_i(t+1)=\sum_{j=1}^{N} J_{ij}\phi(x_j(t)).
\end{equation}
DMF equations can still easily be solved. With discrete-time evolution, the mean field dynamics reads:
\begin{equation}
	x_i(t+1)=\eta_i(t)
\end{equation}
which identifies directly the input current variable $x_i$ with the stochastic process $\eta_i$. In contrast to the continuous case, where self-consistent noise is filtered by a Langevin process, the resulting dynamics is extremely fast.
As a consequence, the statistics of  $\eta_i$ directly translates into the statistics of $x$. We are left with four variables, to be determined according to four equations, which can be synthesized in the following way:
\begin{equation}
	\left(
	\begin{array}{c}
		\mu_E \\
		\mu_I
	\end{array}\right)=J
	\left(
	\begin{array}{cc}
		C_Ej_E  &  -C_Ig_Ej_E\\
		C_Ej_I &  -C_Ig_Ij_I
	\end{array}\right)
	\left(
	\begin{array}{c}
		\left[\phi^E\right]  \\
		\left[\phi^I\right]
	\end{array}\right).
\end{equation}
\begin{equation}
	\left(
	\begin{array}{c}
		\Delta_0^E\\
		\Delta_0^I
	\end{array}\right)=J^2
	\left(
	\begin{array}{cc}
		C_Ej_E^2  &  C_Ig_E^2j_E^2\\
		C_Ej_I^2 &  C_Ig_I^2j_I^2
	\end{array}\right)
	\left(
	\begin{array}{c}
		\left[\phi^{E2}\right]-\left[\phi^E\right]^2  \\
		\left[\phi^{I2}\right]-\left[\phi^I\right]^2
	\end{array}\right).
\end{equation}
As usual, firing rate statistics are computed as averages with respect to a Gaussian distribution with mean $\mu_E$ ($\mu_I$) and variance $\Delta_0^E$ ($\Delta_0^I$).

When adopting discrete-time dynamics, a second condition has to be imposed on the connectivity matrix. To prevent phase-doubling bifurcations specific to discrete-time dynamics, we need the real part of the outliers to be strictly smaller than $r$ in modulus. An isolated outlier on the negative real axis, indeed, would lose stability and induce fast oscillations in the activity before the transition to chaos takes place. The latter condition is satisfied in a regime where inhibition is only weakly dominating, coinciding with the phase region where the approximation provided by DMF is very good (see below in \emph{Methods}).

\paragraph*{Mean field theory with stochastic in-degree}
We derive here the dynamical mean field equations for network in which the total number of inputs $C$ varies randomly between different units in the network.
We focus on a connectivity matrix with one excitatory and one inhibitory column. In the excitatory column, each element $J_{ij}$ is drawn from the following discrete distribution:
\begin{displaymath}
	J_{ij} = \left\{
	\begin{array}{lr}
		J &  p=C_E/N_E=C/N\\
		0 & \mathrm{otherwise}
	\end{array}
	\right.
	\label{Jij}
\end{displaymath} 
Up to the order $O(1/N)$, the statistics of the entries $J_{ij}$ are are:
\begin{equation}\label{eq:stat}
	\langle J_{ij} \rangle=\frac{J}{N}C
\end{equation}
\begin{equation}
	\langle J_{ij}^2\rangle=\frac{J^2}{N}C.
\end{equation}

The inhibitory column is defined in a similar way, if substituting $J$ with $-gJ$.

We proceed in the same order as in the previous sections. We define the effective stochastic coupling, given by $\eta_i(t)=\sum_j J_{ij}\phi(x_j(t))$. We compute the equations for the mean and the correlation of the Gaussian noise $\eta_i$ in the thermodynamic limit. 

We will find that the variance associated to the single neuron activity will consist of a temporal component, coinciding with the amplitude squared of chaotic fluctuations, and of a quenched term, which appears when sampling different realizations of the random connectivity matrix. 

For a given realization and a given unit $i$, the temporal auto-correlation coincides with: $[\eta_i(t)\eta_i(t+\tau)]_{t, ic} - [\eta_i]_{t, ic}^2$ by averaging over time and over different initial conditions. In a second step, averaging over all the units in the population, or equivalently, over the realizations of the matrix $J_{ij}$, returns the average size of deviations from single unit mean within one single trial $[[\eta_i(t)\eta_i(t+\tau)]_{t, ic} - [\eta_i]_{t, ic}^2]_J= [\eta_i(t)\eta_i(t+\tau)] - [[\eta_i]_{t, ic}^2]_J$. In this notation, $[]$ indicates an effective average over time, initial conditions, and matrix realizations. One can compute self-consistently this quantity and check that it coincides with the expression for the total second order moment we found in the previous paragraph for the fixed in-degree case.

In order to close the expression for the DMF equations, we will need to express all the averages of $\phi$ in terms of the total variance $\Delta_0$, which includes quenched variability.
For this reason, we compute the average deviations from $[\eta_i(t)\eta_i(t+\tau)]$ with respect to the population (realizations) mean $[\eta_i]$. As a result, the second moment $[\eta_i(t)\eta_j(t+\tau)]-[\eta_i(t)]^2$ will now include the static trial-to-trial variability.

For the mean, we get:
\begin{equation}
	\begin{split}
		[\eta_i(t)]&=\langle \sum_{j_E=1}^{N_E} J_{ij_E} \phi(x_{j_{E}}(t))\rangle+\langle \sum_{j_I=1}^{N_I} J_{ij_I} \phi(x_{j_{I}}(t))\rangle=\left(N_E\langle J_{ij_{E}}\rangle+N_I\langle J_{ij_{I}}\rangle\right) \langle\phi\rangle\\
		&=J(C_E-gC_I)\langle\phi\rangle.
	\end{split}
\end{equation}

Applying the same steps as before, we compute the second order statistics:
\begin{equation}
	\begin{split}
		&[\eta_i(t)\eta_i(t+\tau)]=\langle\sum_{k=1}^{N} J_{ik} \phi(x_k(t))\sum_{l=1}^{N} J_{il} \phi(x_l(t+\tau))\rangle\\
		&=\langle\sum_{k_E=1}^{N_E} J_{ik_E} \phi (x_{k_E}(t))\sum_{l_E=1}^{N_E} J_{il_E} \phi (x_{l_E}(t+\tau))\rangle+ \langle\sum_{k_I=1}^{N_I} J_{ik_I} \phi (x_{k_I}(t))\sum_{l_I=1}^{N_I} J_{il_I} \phi (x_{l_I}(t+\tau))\rangle\\
		&+ 2 \langle\sum_{k_E=1}^{N_E} J_{ik_E} \phi (x_{k_E}(t))\sum_{l_I=1}^{N_I} J_{il_I} \phi (x_{l_I}(t+\tau))\rangle.
	\end{split}
\end{equation}
Again, we consider separate contributions from diagonal ($k=l$) and off-diagonal ($k \neq l$) terms. This results in:
\begin{equation}
	\begin{split}
		&[\eta_i(t)\eta_i(t+\tau)]=C_E J^2\langle\phi(x_i(t))\phi(x_i(t+\tau))\rangle+C_E^2(1-1/N_E)J^2\langle\phi\rangle^2\\
		&-2C_E C_I gJ^2\langle\phi\rangle^2+C_Ig^2 J^2 \langle\phi(x_i(t))\phi(x_i(t+\tau))\rangle+ C_I^2(1-1/N_I)g^2 J^2 \langle \phi\rangle^2.
	\end{split}
\end{equation}
As we can see, diagonal terms behave, on average, like in the fixed in-degree case. To estimate the off-diagonal contributions, we observe that for every $k_E$ index, the expected number of other non-zero incoming connections is $C_E(1-1/N_E)$. As a consequence, the $k_E\neq l_E$ sum contains on average $C_E^2$ terms of value $J^2\langle \phi\rangle^2$ in the limit $N\rightarrow\infty$. Note that in the fixed in-degree case, the same sum contained exactly $C_E(C_E-1)$ terms. That resulted in a smaller value for the second order statistics, which does not include the contribution from stochasticity in the number of incoming connections.
Similar arguments hold for the inhibitory units.

To conclude, in the large network limit, we found:
\begin{equation}
	\begin{split}
		&[\eta_i(t)\eta_i(t+\tau)] = J^2(C_E+g^2C_I)\langle\phi(x_i(t))\phi(x_i(t+\tau))\rangle + J^2(C_E-gC_I)^2\langle\phi\rangle^2 
	\end{split}
\end{equation}
such that the final result reads:
\begin{equation}
	[\eta_i(t)\eta_i(t+\tau)]-[\eta_i(t)]^2 = J^2(C_E+g^2C_I)C(\tau).
\end{equation}

As before, one can then check that the cross-correlation between different units vanishes.

The noise distribution determines the following self-consistent potential:
\begin{equation}
	V(\Delta, \Delta_0)=-\frac{\Delta^2}{2}+J^2(C_E+g^2 C_I) \int \mathcal{D}z  \left[ \int \mathcal{D}x \Phi (\mu+\sqrt{\Delta_0-\Delta}x+\sqrt{\Delta}z) \right]^2.
\end{equation}

In contrast with the potential of Eq.~\eqref{eq:V_mem}, which was found for networks with fixed in-degree, here we observe the lack of the term $-\Delta[\phi]^2$. As a consequence, the new potential is flat around a non-zero $\Delta=\Delta_\infty$ value, which represents the asymptotic population disorder.

As usually, we derive the DMF solution in the weak and in the strong coupling regime thanks to the analogy with the one-dimensional equation of motion.
When $J<J_C$, the potential has the shape of a concave parabola, the vertex of which is shifted to $\Delta_{\infty} \neq 0$. The only admittable physical solution is here $\Delta(\tau)=\Delta_0=\Delta_{\infty}$.
In order to determine its value, we use the condition emerging from setting $\ddot{\Delta}=0$:
\begin{equation}
	\Delta_0=J^2(C_E+g^2C_I)\int \mathcal{D}z \phi^2(\mu+\sqrt{\Delta_0}z)
\end{equation}
to be solved together with the equation for the mean:
\begin{equation}
	\mu=J(C_E-gC_I)\int \mathcal{D}z \phi(\mu+\sqrt{\Delta_0}z).
	\label{eq:randomc1_mem}
\end{equation}

When $J>J_C$, the auto-correlation acquires a temporal structure. The stable solution is monotonically decreasing from $\Delta_0$ to a value $\Delta_{\infty}$, and we need to self-consistently determine $\mu$, $\Delta_{\infty}$ and $\Delta_0$ through three coupled equations. Apart from the usual one for $\mu$, a second equation is given by the energy conservation law:
\begin{equation}
	V(\Delta_0, \Delta_0)=V(\Delta_{\infty}, \Delta_0)
\end{equation}
which reads:
\begin{equation}
	\begin{split}
		\frac{\Delta_0^2-\Delta_{\infty}^2}{2}=J^2(C_E+g^2C_I)&\left\{\int \mathcal{D}z \Phi^2(\mu+\sqrt{\Delta_0}z) \right. \\
		&\left. -\int \mathcal{D}z  \left[ \int \mathcal{D}y \Phi (\mu+\sqrt{\Delta_0-\Delta_{\infty}}y+\sqrt{\Delta_{\infty}}z) \right]^2 \right\}.
		\label{eq:randomc2_mem}
		\end {split}
	\end{equation}
	The third equation emerges from setting $\ddot{\Delta}=0$ at $\Delta_{\infty}$, which gives:
	\begin{equation}
		\Delta_{\infty}=J^2(C_E+g^2C_I)\int \mathcal{D}z  \left[ \int \mathcal{D}y \phi (\mu+\sqrt{\Delta_0-\Delta_{\infty}}y+\sqrt{\Delta_{\infty}}z) \right]^2.
		\label{eq:randomc3_mem}
	\end{equation}

	\begin{comment}
	The position of the critical coupling $J_C$ can be derived, as in the previous case, thanks to random matrix theory arguments. The stability matrix relative to the distributed fixed point $\{x_{0i}\}$ is given by $S_{ij}=\phi'(x_{0j})J_{ij}$. The radius of its eigenspectrum can be computed as the sum of the variances of the $S_{ij}$ distribution in the excitatory and inhibitory columns, rescaled by $N$:
	\begin{equation}
	r=J^2(C_E+g^2C_I)[\phi'^2(x_{0i})].
	\end{equation}
	The mean in the r.h.s. has to be computed with respect to the gaussian spread in the fixed point solution. The critical coupling is then given by:
	\begin{equation}
	J_C^2(C_E+g^2C_I)\int \mathcal{D}z \phi'^2(\mu+\sqrt{\Delta_0}z)=1.
	\end{equation}*/
	\end{comment}
	
	\subsection*{Finite size effects and limits of the mean field assumptions} 
	
	We test numerically the validity of the Gaussian assumptions and the predictions emerging from the DMF theory.
	We found two main sources of discrepancies between the theory and numerics, namely finite-size effects and the asymmetry between excitation and inhibition.
	
\begin{figure}[h!]
	\begin{adjustwidth}{-0.3in}{-0.3in} 
			\centering
			\includegraphics{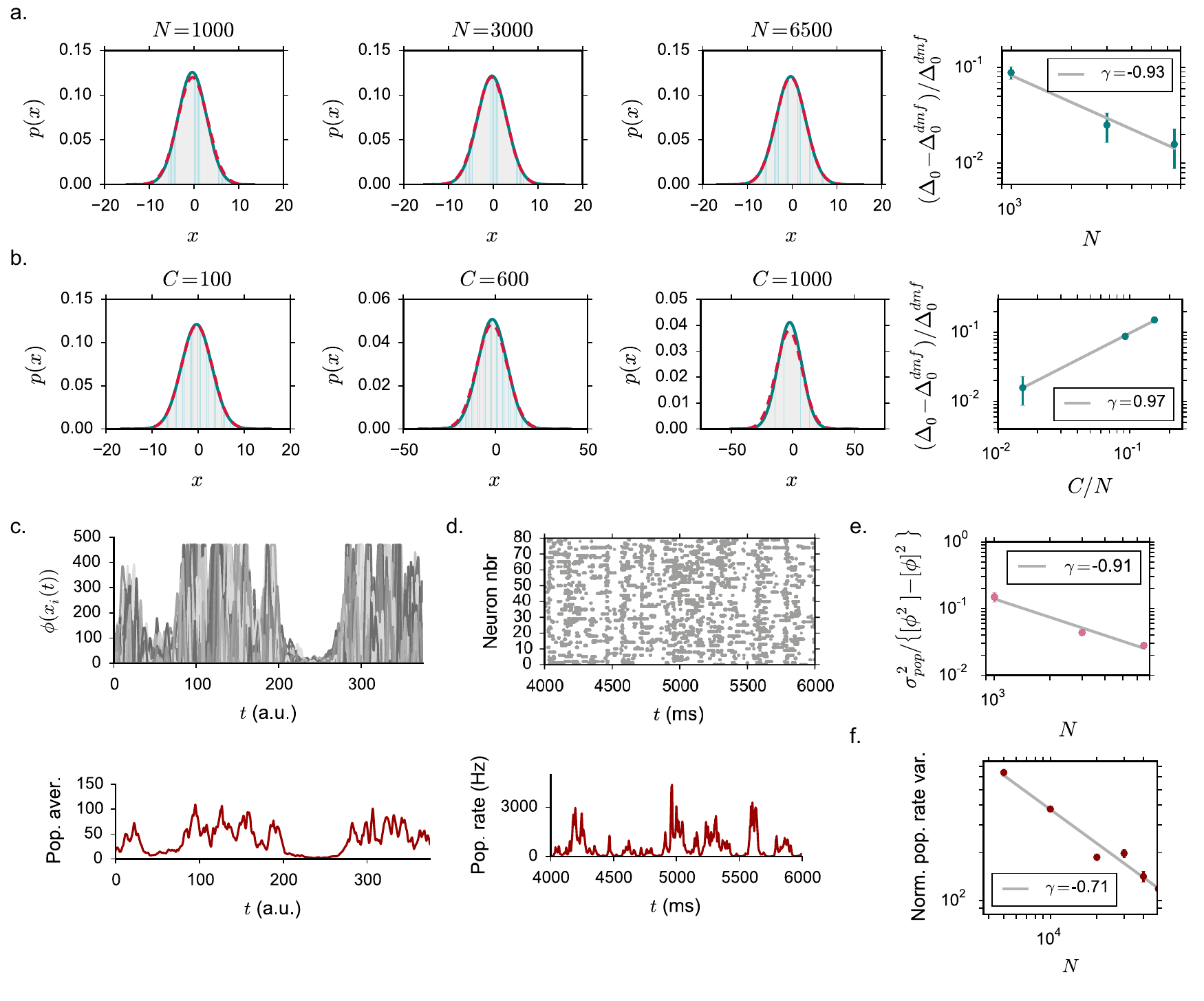}
			\caption{{\bf Comparison between dynamical mean field predictions and numerical simulations: finite size effects.} {\bf a.} Dependence on the system size $N$ ($C=100$). In the first three panels: distribution of the input current $x$ in the population and in different time steps. The numerical distribution is obtained through averaging over 3 realizations of the synaptic matrix. Light green: simulated data distribution, dark green: best Gaussian fit to data, red: DMF prediction. In the fourth panel: normalized deviations from the DMF theoretical value. The log-log dependence is fitted with a linear function, $\gamma$ giving the coefficient of the linear term. Choice of the parameters: $g=4.1$, $J=0.2$, $\phi_{max}=2$. 
				{\bf b.} As in {\bf a}, dependence on the in-degree $C$ ($N=6500$). 
				{\bf c.} Finite size effects in rate networks with large saturation upper-bound: sample of network activity (top: single units in grey scale, bottom: population averaged firing rate). Choice of the parameters: $g=5$, $J=0.14$, $\phi_{max}=240$. 
				{\bf d.} Finite size effects in networks of LIF neurons with small refractory period: sample of network activity (rastergram of 80 randomly selected neurons, population averaged firing rate). Choice of the parameters: $N=20000$, $C=500$, $g=5$, $\tau_{rp}=0.01$ ms, $J=0.9$ mV.
				{\bf e.} Finite size effects in rate networks with large saturation upper-bound: 
				normalized variance of the  population-averaged firing rate as a function of the network size. 
				{\bf f.} Finite size effects in networks of LIF neurons with small refractory period: normalized variance of the  population-averaged firing rate as a function of the network size (computed with 1 ms bins).}
			\label{fig:fs}
		\end{adjustwidth}
	\end{figure}

	As a first step, we analyzed the magnitude of finite size effects deriving from taking finite network sizes.
	Fig.~\ref{fig:fs} {\bf a} shows a good agreement between simulated data and theoretical expectations. The magnitude of finite size effects shrinks as the network size is increased and cross-correlations between different units decays. 
	
	In Fig.~\ref{fig:fs} {\bf b}  we tested instead the effect of increasing the in-degree $C$ when $N$ is kept fixed. When $C$ is constant and homogeneous in the two populations, our mean field approach requires network sparseness ($C\ll N$). Consistently, we find an increase in the deviations from the theoretical prediction when $C$ is increased. 
	
	Both the $N$ and $C$ dependencies have the effect of weakly reducing fluctuations variance with respect to the one expected in the thermodynamic limit.
	The numerically obtained $x$ distribution is in good agreement with the assumption of DMF, which states that current variables $x_i$ are distributed, for large time $t$ and size $N$, according to a Gaussian distribution of mean $\mu$ and variance $\Delta_0$.
	
	We observe that stronger deviations from the theoretical predictions can arise when the upper-bound $\phi_{max}$ on the transfer function is large and the network is in the intermediate and strong coupling regime.
	By simulating the network activity in that case, we observe stronger cross-correlations among units, which can cause larger fluctuations in the population-averaged firing rate.
	
	In Fig.~\ref{fig:fs} {\bf c} we check that those deviations can still be understood as finite size effects: the distance between the DMF value and the observed ones, which now is larger, decreases with $N$ as the correlation among units decay. Equivalently, the variance of the fluctuations in the population-averaged input current and firing rate decays consistently as $\sim 1/N$.
	
	The same effect, and even stronger deviations, are observed in rate models where the transfer function is chosen to mimic LIF neurons.
	
	As a side note, we remark that strong correlations in numerical simulations are observed also in the case of spiking networks of LIF neurons with small refractory period and intermediate coupling values (Fig.~\ref{fig:fs} {\bf d}) .
	Also in this case, correlations are reflected in strong time fluctuations in the population averaged firing rate. Their amplitude should scale with the system size as $1/N$ in the case of independent Poisson processes. This relationship, which is well fitted in the weak and strong coupling regimes, appears to transform into a weaker power law decay for intermediate $J$ values.

\begin{figure}[h!]
	\begin{adjustwidth}{-0.3in}{-0.3in} 
			\centering
			\includegraphics{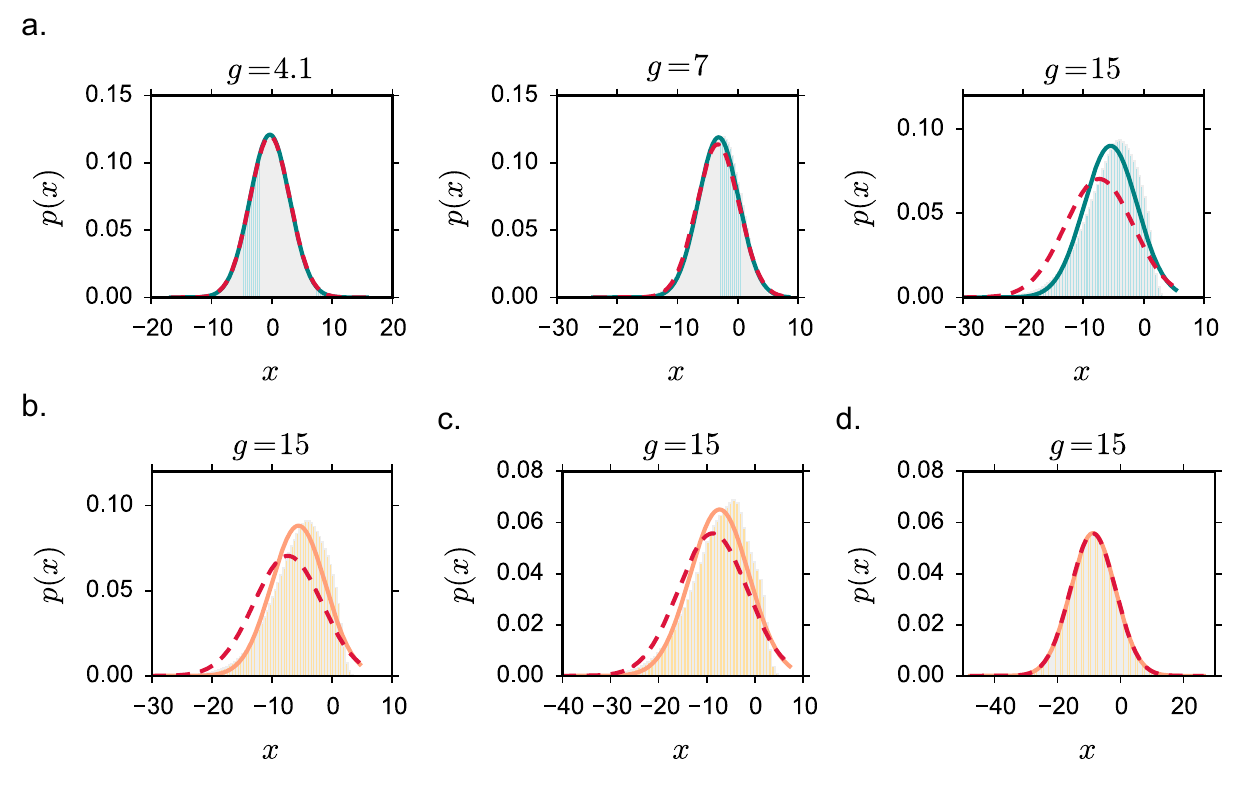}
			\caption{{\bf Comparison between dynamical mean field predictions and numerical simulations: the effects of strong inhibition.} Distribution of the input current $x$ in the population and in different time steps. The numerical distribution is obtained through averaging over 3 realizations of the synaptic matrix. Light green/orange: simulated data distribution, dark green/orange: best Gaussian fit to data, red: DMF prediction. Choice of the parameters: $C=100$, $N=6500$, $J=0.2$. {\bf a.} Dependence on the inhibition dominance $g$. {\bf b.} Numerical distribution for a network with a synaptic matrix where $C$ is fixed, as above, but excitatory and inhibitory units are shuffled. {\bf c.} As above, with a synaptic matrix where $C$ is random. {\bf d.} As above, with the equivalent Gaussian matrix, whose statistics match the ones of the sparse one. }
			\label{fig:depg}
		\end{adjustwidth}
	\end{figure}

	\paragraph{Limits of the Gaussian approximation}
	A different effect is found by increasing the dominance of inhibition over excitation  in the network, i.e. by increasing $g$, or equivalently, by decreasing $f$. As shown in Fig.~\ref{fig:depg} {\bf a} , inhibition dominance can significantly deform the shape of the distribution, which displays suppressed tails for positive currents. As the inhibition dominance is increased, since $\phi(x_i)$ is positive and $J_{ij}$ strongly negative on average, the fluctuations become increasingly skewed in the negative direction.
	As expected, the Gaussian approximation does not fit well the simulated data. Fig.~\ref{fig:fs} {\bf b-c-d}  shows that the same effect is quite general and extends to networks where excitation and inhibition are not segregated or the connectivity $C$ is random.
	
	An extreme consequence of this effect is the failure of DMF in describing purely inhibitory networks in absence  of external excitatory currents, where the effective coupling $\eta_i(t)=\sum_{j}J_{ij}\phi(x_j(t))$ is strictly non-positive at all times. In this case, DMF erroneously predicts a critical coupling $J_D$ between a bounded  and an unbounded regime, the divergence being led by the positive tails of the Gaussian bell. In contrast, in absence of any positive feedback, purely inhibitory networks cannot display a transition to run-away activity.
	
	As a final remark, we observe that the agreement between simulated activity and mean field predictions in the case of purely inhibitory networks is in general less good than the one we found for EI architectures.
	
	We conclude that the Gaussian hypothesis adopted in the DMF framework is a reasonable approximation only when inhibition does not overly dominate excitation.
	Finally, we remark that this limitations critically depends on adopting sparse matrices where  non-zero entries have fixed values. If adopting a Gaussian, fully-connected connectivity, whose mean and variance are matching the ones of the original matrix:
	\begin{equation}
		\begin{split}
			&[J_{ij}]=\frac{J}{N}(C_E-gC_I)\\
			&[J_{ij}^2]=\frac{J^2}{N}(C_E+g^2C_I)
		\end{split}
	\end{equation}
	numerical simulations reveal that, whatever the degree of inhibition, positive entries are strong enough to balance the distribution, which strongly resembles again a Gaussian bell. 
	
	\subsection*{Network of integrate-and-fire neurons}
	
	The simulations  presented in  Fig.~\ref{fig:spk} were performed  on a
	network   of    leaky   integrate-and-fire (LIF)  neurons    identical   to
	\cite{Ostojic}. The  membrane
	potential dynamics of the $i$-th LIF neuron are given by:
	
	\begin{equation}
		\tau_m\frac{dV_i}{dt}=-V_i+\mu_0 + RI_i(t)+\mu_{\mathrm{ext}}(t)
	\end{equation}
	where  $\tau_m=20$ ms  is the  membrane  time constant,  $\mu_0$ is  a
	constant  offset current, and $RI_i$  is the  total  synaptic input  from
	within the  network. When the membrane potential crosses the threshold $V_{\mathrm{th}}=20$
	mV, an action potential is emitted and the membrane potential is reset
	to  the value  $V_{\mathrm{r}}=10$ mV.  The  dynamics resume  after a  refractory
	period  $\tau_{\mathrm{r}}$, the value of which was systematically varied.  The total synaptic input to the i-th neuron is:
	\begin{eqnarray}
		RI_i(t)=\tau_m \sum_j J_{ij}\sum_k \delta(t-t_j^{(k)}-\Delta)
	\end{eqnarray}
	where  $J_{ij}$ is  the amplitude of the post-synaptic potential evoked in neuron $i$ by an action potential occurring in neuron $j$, and $\Delta$ is the synaptic delay (here taken to be 1.1 ms).  Note that if the synaptic delay is shorter than the refractory period, the network develops spurious synchronization \cite{Ostojic2015}.
	
	The connectivity matrix $J_{ij}$ was identical to the rate network with fixed in-degree described above.

	\section*{Acknowledgements}
	We are grateful to Vincent Hakim and Nicolas Brunel for discussions and feedback on the manuscript.
	
	%\section*{References}
	% Either type in your references using
	% \begin{thebibliography}{}
	% Text
	% \end{thebibliography}
	%
	% OR
	%
	% Compile your BiBTeX database using our plos2015.bst
	% style file and paste the contents of your .bbl file
	% here.
	% 

\end{document}